\documentclass[%
 reprint,
showpacs,
 amsmath,amssymb,
 aps,
 pra,
floatfix,
superscriptaddress
]{revtex4-1}
\usepackage[utf8]{inputenc}
\usepackage[T1]{fontenc}
\usepackage[]{graphicx}
\usepackage{color}
\usepackage[dvipsnames]{xcolor}
\usepackage[english]{babel}
\usepackage{ulem}
\usepackage[bookmarks,bookmarksopen,bookmarksdepth=2]{hyperref}
\hypersetup{
    colorlinks=true,
    citecolor=blue,
    linkcolor=blue,
    filecolor=magenta,
    urlcolor=cyan,
}

\usepackage{url}

\usepackage{microtype}
\usepackage{braket}
\usepackage{esvect}
\usepackage[export]{adjustbox}

\usepackage{amsmath,amsfonts,amssymb}
\usepackage{mathtools}
\usepackage{bbm}

\usepackage[position=top,caption=false]{subfig}

\let\originaleqref\eqref
\renewcommand{\eqref}{Eq.~\originaleqref}

\newcommand{\fref}[1]{Fig.~\ref{#1}}

\newcommand{\etal}{\textit{et al. }}

\graphicspath{
{figures/}
}

\begin{document}
\title{Probing scrambling and operator size distributions using random mixed states and local measurements}

\author{Philip Daniel Blocher}
\email{blocher@unm.edu}
\affiliation{Center for Quantum Information and Control, Department of Physics and Astronomy, University of New Mexico, Albuquerque, New Mexico 87131, USA}
\author{Karthik Chinni}
\email{karthik.chinni@polymtl.ca}
\affiliation{Department of Engineering Physics, \'Ecole Polytechnique de Montr\'eal, 2500 Chem. de Polytechnique, Montr\'eal, Quebec H3T 1J4, Canada}
\affiliation{Center for Quantum Information and Control, Department of Physics and Astronomy, University of New Mexico, Albuquerque, New Mexico 87131, USA}
\author{Sivaprasad Omanakuttan}
\email{somanakuttan@unm.edu}
\affiliation{Center for Quantum Information and Control, Department of Physics and Astronomy, University of New Mexico, Albuquerque, New Mexico 87131, USA}
\author{Pablo M. Poggi}
\email{pablo.poggi@strath.ac.uk}
\affiliation{Department of Physics, SUPA and University of Strathclyde, Glasgow G4 0NG, United Kingdom}
\affiliation{Center for Quantum Information and Control, Department of Physics and Astronomy, University of New Mexico, Albuquerque, New Mexico 87131, USA}

\date{\today}
\bigskip

\begin{abstract} 
The dynamical spreading of quantum information through a many-body system, typically called scrambling, is a complex process that has proven to be essential to describe many properties of out-of-equilibrium quantum systems.
Scrambling can, in principle, be fully characterized via the use of out-of-time-ordered correlation functions, which are notoriously hard to access experimentally. 
In this work, we put forward an alternative toolbox of measurement protocols to experimentally probe scrambling by accessing properties of the operator size probability distribution, which tracks the size of the support of observables in a many-body system over time. Our measurement protocols require the preparation of separable mixed states together with local operations and measurements, and combine the tools of randomized operations, a modern development of near-term quantum algorithms, with the use of mixed states, a standard tool in NMR experiments.
We demonstrate how to efficiently probe the probability-generating function of the operator distribution and discuss the challenges associated with obtaining the moments of the operator distribution.
We further show that manipulating the initial state of the protocol allows us to directly obtain the individual elements of the distribution for small system sizes.
\end{abstract}

\maketitle
\noindent

\section{Introduction}\label{section:introduction}
Understanding how quantum information spreads across the degrees of freedom of a quantum system is a key part of developing a comprehensive picture of nonequilibrium quantum many-body physics. In this context, the notion of scrambling has attracted much attention over the past years due to its relevance in the study of closed-system thermalization \cite{NatCommun.10.1581}, quantum chaos \cite{fortes2019}, information retrieval in black holes \cite{Yasuhiro_Sekino_2008,Patrick_Hayden_2007}, and quantum algorithms \cite{dupont2022}. Scrambling refers to the dynamical delocalization of quantum information \cite{zhuang2019scrambling} and can be diagnosed by the generation of entangled states from initially separable ones, or from the growth of initially local operators \cite{Parker2019}. 

An approach to characterizing scrambling from the unitary evolution of an operator $W(t)$ in a many-body system is to analyze the dynamics of so-called operator size distributions $\{P_k(t)\}$ \cite{Roberts2018,schuster2022,Omanakuttan2023scrambling}. These can be obtained by coarse-graining the expansion of $W(t)$ in a complete operator basis, the choice of which depends on the nature of the system. In the case of systems of $N$ spin$-\frac{1}{2}$ particles, a natural operator basis is the set of multi-body Pauli operators $\mathcal{P}_N=\{\mathbbm{1},\sigma_x,\sigma_y,\sigma_z\}^{\otimes N}/\sqrt{2^N}$, which has dimension $D=4^N$ and forms an orthonormal basis. In the Heisenberg picture an operator $W$ may at time $t$ be written as
\begin{equation}
W(t) = \mathcal{U}^\dagger(t)\, W\, \mathcal{U}(t) = \sum_{j=0}^{D-1} f[\Lambda_j;\, W(t)] \Lambda_j, \label{eq:operatorbasisexpansion}
\end{equation}
where $\mathcal{U}(t)$ is the unitary time evolution operator from the initial time $t=0$ to time $t$ and $\Lambda_j\in\mathcal{P}_N$. Given this expansion of $W(t)$, the operator size distribution is constructed by grouping the elements of the exponentially large Pauli basis according to their size $s(\Lambda)$. Here the operator size corresponds to the number of non-identity operators in the Pauli string (i.e., its Hamming weight) and thus $1\leq s(\Lambda)\leq N$ such that $\mathcal{P}_N = \cup_{k=1}^N C_k$, where $C_k = \{ \Lambda\, \vert\, s(\Lambda) = k \}$. The resulting operator size distribution reads
\begin{equation}
    P_k(t) = \frac{1}{\text{Tr}[W^\dagger W]} \sum_{\Lambda\in C_k} \vert f[\Lambda;\,W(t)] \vert^2, \label{eq:coarsegrainedprobabilitydist}
\end{equation}
and measures the size of the support of $W(t)$. It is easy to see that $\sum_{k=1}^N P_k(t) = 1$ for all times $t$, hence the operator size distribution can be regarded as a coarse-grained probability distribution in the expansion coefficients of $W(t)$.

Of particular interest is the case where $W(0)$ is a size-one (i.e. single-body) Pauli operator such that $P_k(0)=\delta_{k,1}$. As the operator grows and information becomes scrambled, the distribution shifts to higher values of $k$ and grows in variance. The dynamics of $P_k(t)$ have been studied for various many-body models \cite{Omanakuttan2023scrambling,Roberts2018}, and it holds a close connection to the Krylov picture of operator growth \cite{Parker2019,noh2021}.

\begin{figure}[t!]
    \centering
    \includegraphics[width=0.9\linewidth]{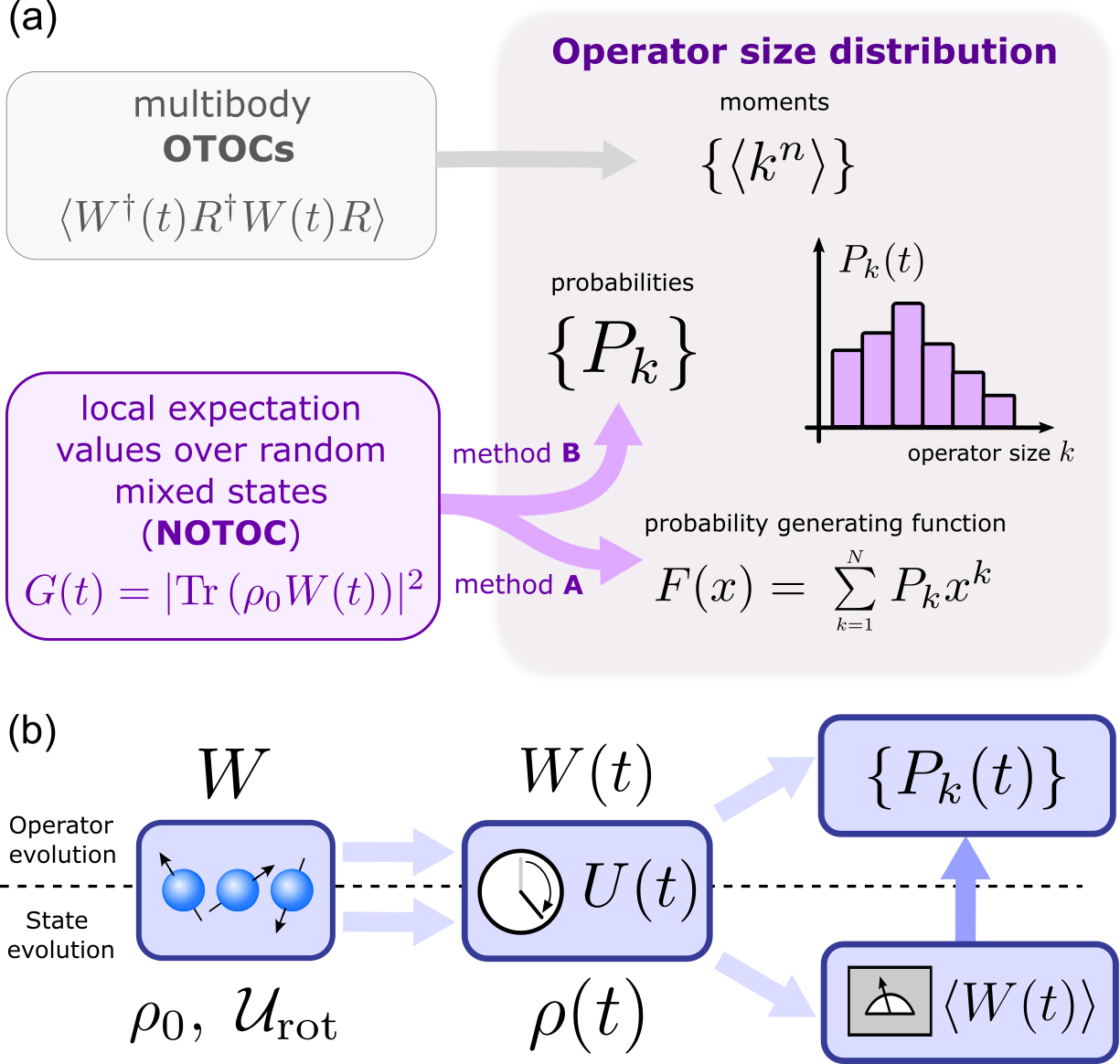}
    \caption{(a) Relationship between the operator size distribution $P_k(t)$ of \eqref{eq:coarsegrainedprobabilitydist}, out-of-time-ordered correlation functions (OTOCs), and the NOTOC approach proposed in this article using local expectation values over random mixed states. OTOCs allow (indirect) access of the moments of the operator distribution, while our proposal allows to probe the probability generating function $F(x)$ or the individual probabilities directly, depending on the choice of initial state $\rho_0$. (b) Illustration of the operator evolution (above the dashed line) and the proposed measurement protocol (below the dashed line). The operator size distribution $P_k(t)$ is a property of an operator in the Heisenberg picture (and thus independent of a choice of state). Our protocol proposes to access this property by measuring expectation values $\braket{W(t)} = \text{Tr}[\rho(t) W]$ after the time evolution of a suitably-chosen set of initial states $\rho_0$.}
    \label{fig:PD-PFG-OTOC-comparison}
\end{figure}

While the scrambling dynamics in the system may be described using the operator size distribution, it is not obvious how to access this distribution experimentally. A complete characterization of $P_k(t)$ for all $k$ requires an exponential amount of resources; this is clearly seen if one reconstructs $P_k(t)$ from two-point correlation functions to obtain each of the coefficients $f[\Lambda;W(t)]=\text{Tr}\left(\Lambda W(t)\right)$ in \eqref{eq:operatorbasisexpansion}. A partial workaround is given by considering out-of-time-ordered correlators (OTOCs) which take the form
\begin{equation}
    \mathcal{F}\left(W(t),R\right) = \braket{W^\dagger(t) R^\dagger(0) W(t) R(0)}. \label{eq:standardOTOC}
\end{equation}
It has been shown that moments of $P_k(t)$ can be computed by constructing averages of OTOCs over appropriate sets of operators $\{R_i\}$ \cite{Roberts2018,qi2019,hosur2016,Omanakuttan2023scrambling}. However, accessing even a single OTOC is often challenging in experiments due to the out-of-time-ordered nature of Eq. (\ref{eq:standardOTOC}), and typically requires the use of many-body time-reversal operations \cite{Garttner2017,PhysRevLett.120.040402,PhysRevX.7.031011,PhysRevLett.123.090605,arXiv.1607.01801} or auxiliary systems \cite{nature.567.61}. In cases where OTOCs can be accessed without these tools (see for instance the method in \cite{PhysRevX.9.021061}), reconstructing the moments of $P_k(t)$ requires the measurement of a large number of OTOCs and the task becomes unfeasible for high-order moments~\cite{Omanakuttan2023scrambling}.

In this article we propose an alternative set of tools to experimentally probe the operator size distribution which circumvents the use of OTOCs completely. We combine the use of ensembles of initially separable mixed states with local random operations and local measurements at the final time in order to access a quantity ${G(t)}$ which is explicitly \textit{not} an OTOC, hence we name this quantity a ``NOTOC''. We show that our measurement protocol, depending on the choice of initial state ensemble, probes either the probability generating function (PGF) $F(x,t)=\sum_k P_k(t) x^k$ of the operator size distribution (method A), or its elements $\{P_k(t)\}$ directly (method B), as illustrated in \fref{fig:PD-PFG-OTOC-comparison}. We demonstrate our methods numerically and show that the operator size distribution can be accurately probed even when accounting for statistical noise stemming from averaging over random operations and from experimental imperfections. In our analysis of method A, we discuss inherent shortcomings of the problem of inverting the PGF to obtain $P_k(t)$ such as its sensitivity to noise, and we show for the case of the Ising model that the PGF can itself be seen as a good indicator of the presence of scrambling in the system. For method B we discuss the efficiency of the method as the system size $N$ increases and show that individual probabilities $P_k(t)$ can be reliably obtained as long as $k\ll N$.

Our proposed measurement protocol connects to previous works which focused on experimental schemes to diagnose scrambling. In particular, method A recovers a procedure put forward in Ref.~\cite{qi2019} to measure operator growth in the special case when the initial states are pure. Likewise, the use of randomized operations to access properties of the operator size distribution makes this proposal complementary to the one in Ref.~\cite{PhysRevX.9.021061}, where similar tools were used to measure OTOCs instead. Finally, the NOTOC proposed here can be seen as a generalization of the fidelity OTOC \cite{NatCommun.10.1581,Garttner2017,PhysRevLett.120.040402}, which has been widely studied due to its ease of accessibility in experiments \cite{PhysRevA.106.042429}. Our analysis reveals that this quantity is inherently connected to the operator size distribution, although in a way that is fundamentally different from regular OTOCs, and that had not been revealed up to now to the best of our knowledge.

The structure of this article is as follows: In Sec.~\ref{section:protocol} we present the main tools to be used to probe the operator size distribution $P_k(t)$, including the choice of initial states, randomized operations and local measurements. Sections~\ref{section:epsilonprotocol} and \ref{section:nonepsilonprotocol} discuss two methods of implementing the necessary state preparation for our measurement protocol, with the method of Sec.~\ref{section:epsilonprotocol} using partially polarized qubit states to obtain the probability-generating function for the operator size distribution, and the method of Sec.~\ref{section:nonepsilonprotocol} tailoring separable states to obtain the probability distribution elements directly. In Sec. \ref{section:discussion} we compare our measurement protocol to previous results and proposals in the literature, extend the NOTOC measurement protocol to collective spin systems, and provide a general discussion of the NOTOC toolbox. Finally, we conclude on our work in Sec.~\ref{section:Outlook} and discuss possible future extensions.

\section{NOTOC measurement protocol}\label{section:protocol}
Our measurement protocol draws inspiration from the so-called fidelity OTOCs, which are OTOCs on the form \eqref{eq:standardOTOC} where one chooses the operator $R = \rho_0$ in \eqref{eq:standardOTOC} to be the projector onto the pure initial state $\rho_0$ \cite{NatCommun.10.1581,Garttner2017,PhysRevLett.120.040402}. Fidelity OTOCs are experimentally accessible quantities as \cite{PhysRevA.106.042429}
\begin{equation}
    \braket{W^\dagger(t) R^\dagger(0) W(t) R(0)} \equiv \vert \braket{W(t)}\vert^2, \label{eq:FOTOCidentity}
\end{equation}
hence they require only the measurement of a single-time expectation value. However, this identity holds only for pure initial states, and the fidelity OTOCs for mixed initial states takes on a more elaborate form, as discussed in Ref.~\cite{PhysRevA.106.042429}.

Inspired by the simple form of the fidelity OTOC in \eqref{eq:FOTOCidentity} as a squared expectation value of an operator $W$ for a pure state, we define the quantity of interest for our protocol as
\begin{equation}
    G(t) := \vert \braket{W(t)}\vert^2 = \vert \text{Tr}[\rho_0 W(t)]\vert^2, \label{eq:NOTOCdefinition}
\end{equation}
with $\rho_0$ being a generic mixed state. The quantity $G(t)$ defined in \eqref{eq:NOTOCdefinition} is equivalent to an OTOC only for pure states \footnote{For pure states Eq.~\ref{eq:NOTOCdefinition} is equivalent to a fidelity OTOC, which in turn is a special case of the more general OTOC in Eq.~(\ref{eq:standardOTOC}) when one chooses $R(0) = \rho_0 = \ket{\Psi_0}\bra{\Psi_0}$.}, and to emphasize that $G(t)$ is related to fidelity OTOCs while itself being a time-ordered correlation function at all times and for all initial states,  we will refer to $G(t)$ as a ``NOTOC'' throughout this article. In the remainder of this section we will demonstrate the utility of defining $G(t)$ on the form \eqref{eq:NOTOCdefinition}, and in Sec.~\ref{section:discussion} we will comment further on its relation to fidelity OTOCs.

Using \eqref{eq:operatorbasisexpansion} the NOTOC $G(t)$ may be written on the form
\begin{align}
    G(t) =& \sum_{i} \vert f[\Lambda_i;\, W(t)]\vert^2 \braket{\Lambda_i}^2 \nonumber\\
    &+ \sum_{i\neq j} f[\Lambda_i;\, W(t)] f[\Lambda_j;\, W(t)]^\ast \braket{\Lambda_i}\braket{\Lambda_j}^\ast. \label{eq:protoNOTOC}
\end{align}
We now consider a generic mixed state on the form
\begin{equation}
\rho_0 = \frac{1}{d}\mathbbm{1} + \sum_i r_i \Lambda_i, \label{eq:initialstateexpansion}
\end{equation}
for which $\braket{\Lambda_i} = r_i$. We recall from Sec.~\ref{section:introduction} that $C_k = \{ \Lambda\, \vert\, s(\Lambda) = k \}$ is the set of Pauli operators of size $k$. If we were to engineer an ensemble of initial states $\{\rho_0\}$ on the form \eqref{eq:initialstateexpansion} such that the state coefficients $\{r_i\}$ were independent, identically distributed random variables with vanishing mean $\overline{r_i}=0$, the second term of \eqref{eq:protoNOTOC} would vanish under averaging over this ensemble. Furthermore, if we require the state coefficients of the engineered initial state to have finite variance $\overline{r_i^2} = \Delta_k$ for $\Lambda_i \in C_k$, the averaged quantity then reads
\begin{align}
    \overline{G(t)} =& \sum_{i} \vert f[\Lambda_i;\, W(t)]\vert^2\, \overline{r_i^2} \nonumber\\
    &+ \sum_{i\neq j} f[\Lambda_i;\, W(t)] f[\Lambda_j;\, W(t)]^\ast\, \overline{r_i r_j^\ast} \\
    =& \sum_{k=1}^N\sum_{\Lambda_i\in C_k} \vert f[\Lambda_i;\, W(t)]\vert^2 \Delta_k \label{eq:NOTOCp} \\
    =& \sum_{k=1}^N \Delta_k \text{Tr}[W^\dagger W] P_k(t). \label{eq:NOTOC}
\end{align}
Equation~(\ref{eq:NOTOC}) reveals that $\overline{G(t)}$ is a linear combination of the elements of the probability distribution $\{P_k(t)\}$, with coefficients proportional to the variance $\Delta_k$ times the 2-norm $\text{Tr}[W^\dagger W]=\Vert W \Vert_2^2$ of the operator $W$. The probability distribution $\{P_k(t)\}$ may be extracted from \eqref{eq:NOTOC} using several methods, and we present two different measurement protocols for systems of spin-$1/2$ particles in the following sections~\ref{section:epsilonprotocol} and \ref{section:nonepsilonprotocol}. Our measurement protocols provide experimental access to the averaged NOTOC $\overline{G(t)}$ and the probability distribution $\{P_k\}$ through engineering of the initial state $\rho_0$ and subsequent measurement of the expectation value $\braket{W(t)} = \text{Tr}[\rho(t) W]$ at the final time~$t$. The main challenge thus lies in the preparation of random initial states $\rho_0$ whose coefficients $\{r_i\}$ must have appropriate statistics. 

The operator size distribution is an operator property independent of the initial state of the system, hence the Heisenberg picture lends itself nicely to the analysis of the operator size distribution's evolution in time. However, in the NOTOC measurement protocol outlined above we probe the operator size distribution using expectation values $\braket{W(t)}_{\rho_0}\equiv \text{Tr}[\rho_0 W(t)]$, which depend on the choice of initial state $\rho_0$. In this way, the initial state is a control knob used by this protocol to access properties of the operator. As expectation values are quantities independent of the choice of picture, it is more natural to describe our proposed experimental measurement protocol in the Schr{\"o}dinger picture as the time evolution of an initial state $\rho_0 \rightarrow \rho(t)$, with which we first evaluate the expectation value $\braket{W(t)}_{\rho_0}\equiv \text{Tr}[\rho(t) W]$, subsequently calculate the NOTOC \eqref{eq:NOTOCdefinition}, and finally recover the averaged NOTOC \eqref{eq:NOTOC} by appropriate averaging over initial states. The NOTOC measurement protocol is thus illustrated in this way in \fref{fig:PD-PFG-OTOC-comparison}(b), where the Heisenberg operator evolution is shown above the dashed line, and our proposed measurement protocol is illustrated below the dashed line.

\section{Method A: accessing the probability generating function of the probability distribution}\label{section:epsilonprotocol}
In this section we present an experimentally relevant measurement protocol for obtaining the squared expectation value \eqref{eq:NOTOC} using mixed states similar to those used in NMR together with random local operations~\cite{arXiv.2209.09322}. 
The starting point of our measurement protocol is the preparation of the product state
\begin{align}
    \rho_\text{ini} =& \left(\frac{\mathbbm{1}+\varepsilon\sigma_z}{2}\right)^{\otimes N} \\
    =& \frac{1}{d}\left(\mathbbm{1} + \varepsilon \sum_{i} \sigma_z^i + \varepsilon^2 \sum_{i<j} \sigma_z^i \sigma_z^j + \ldots \right) \\
    =& \frac{1}{d}\mathbbm{1} + \sum_{k=1}^{N} \frac{\varepsilon^k}{\sqrt{d}} \sum_{\Lambda \in C_k^z} \Lambda, \label{eq:polarizedqubitstate}
\end{align}
where each qubit is in a statistical mixture of being maximally mixed $\mathbbm{1}$ and polarized along the $z$-axis, with the parameter $\varepsilon$ controlling the amount of polarization. In \eqref{eq:polarizedqubitstate} we have defined the subset $C_k^z \subset C_k$ of size-$k$ Pauli operators that consist of only $\sigma_z$ terms (e.g., $\mathbbm{1}\otimes\sigma_z\otimes\mathbbm{1}$). 
The polarization parameter $\varepsilon$ takes values $\vert\varepsilon\vert\leq1$, and we note that for $\varepsilon=\pm1$ the initial state $\rho_\text{ini}$ is a pure state.

To create a random state $\rho_0$ for the experiment whose expansion coefficients $r_i$ in \eqref{eq:initialstateexpansion} satisfy the appropriate statistics for \eqref{eq:NOTOC}, we apply random local rotations to the initial state \eqref{eq:polarizedqubitstate} via the unitary $\mathcal{U}_\text{rot} = \bigotimes_{i=1}^N \mathcal{U}_\text{rot}^{(i)}$
\begin{align}
    \rho_0 =& \mathcal{U}_\text{rot}\, \rho_\text{ini}\, \mathcal{U}_\text{rot}^\dagger \\
    =& \frac{1}{d}\mathbbm{1} + \sum_{k=1}^N \frac{\varepsilon^k}{\sqrt{d}} \left( \mathcal{U}_\text{rot}\sum_{\Lambda\in C_k^z}  \Lambda\, \mathcal{U}_\text{rot}^\dagger\right) \\
    =& \frac{1}{d}\mathbbm{1} + \sum_{k=1}^N \frac{\varepsilon^k}{\sqrt{d}} \sum_{Q\in C_k} q_{Q}\, Q, \label{eq:initialstatepauliexpansion}
\end{align}
where $\mathcal{U}_\text{rot}^{(i)}$ is a random rotation of the $i$th qubit that we will discuss momentarily. In the last equality of \eqref{eq:initialstatepauliexpansion} we expanded $\mathcal{U}_\text{rot} \sum_{\Lambda\in C^z_k}\, \Lambda\, \mathcal{U}_\text{rot}^\dagger$ on the Pauli operators $Q\in C_k$, as the random local rotations do not change the operator size. Comparing the form of \eqref{eq:initialstatepauliexpansion} to that of \eqref{eq:initialstateexpansion}, we make the identification $r_Q = q_Q \varepsilon^k/\sqrt{d}$ for $s(Q) = k$. We choose the random local rotations $\mathcal{U}_\text{rot}$ such that the coefficients $\{q_Q\}$ are independent, identically distributed random variables with vanishing mean $\overline{q_Q}=0$ and finite variance $\overline{q_Q^2} = \Delta_k$ for $s(Q) = k$. This yields $\overline{r_Q} = 0$ and $\overline{r_Q^2} = \varepsilon^{2k} \Delta_k/d$, which is consistent with the assumptions made in Sec.~\ref{section:protocol}. Substituting this back into \eqref{eq:NOTOC} -- and choosing $W$ to be a non-identity observable with trace $\text{Tr}[W^2] = d$ \footnote{This assumption still allows us to choose $W$ to be any unnormalized Pauli operator from the set $\{\mathbbm{1},\sigma_x,\sigma_y,\sigma_z\}^{\otimes N}$, which would be the natural operators to measure for an $N$-qubit system.} -- we thus find averaged squared expectation value
\begin{equation}
    \overline{G(\varepsilon,t)} = \sum_{k=1}^N P_k(t)\, \Delta_k\, \varepsilon^{2k}. \label{eq:AveragedNOTOC}
\end{equation}
Equation~(\ref{eq:AveragedNOTOC}) is an (at most) $N$th degree polynomial in $\varepsilon^2$ with coefficients proportional to elements $P_k(t)$ of the probability distribution of interest. This form is reminiscent of a probability-generating function (PGF) of the probability distribution $\{P_k(t)\}$ \cite{UnivariateDiscreteDistributions}, and we now show that \eqref{eq:AveragedNOTOC} is indeed a PGF by introducing the explicit form of $\Delta_k$.

To obtain the correct statistics for the coefficients $r_Q$, as well as to ensure that all operators $Q\in C_k$ are sampled for all $k$, we propose to take the single-qubit rotation operator $\mathcal{U}_{\mathrm{rot}}^{(i)}$ to be sampled from a uniform distribution over SU$(2)$. In each random instance the local rotation transforms the $i$th site Pauli-Z as 
\begin{equation}
    \sigma_z^i \rightarrow \mathcal{U}_{\mathrm{rot}}^{(i)} \sigma_z^i \mathcal{U}_{\mathrm{rot}}^{(i)\dagger} = \sum_{\alpha} n_{\alpha}^{(i)} \sigma_{\alpha}^i
\end{equation}
with $\alpha=x,y,z$, with
\begin{align}
    \mathbf{n}^{(i)} =& \cos(\phi_i)\sin(\theta_i) \mathbf{\hat{x}} + \sin(\phi_i)\sin(\theta_i)\mathbf{\hat{y}} \nonumber\\
    &+ \cos(\theta_i) \mathbf{\hat{z}}.
\end{align}
The polar angle $\theta_i$ and azimuthal angle $\phi_i$ are thus random variables taking values in the interval $[0,\pi)$ and $[0, 2\pi)$, respectively.

The coefficients $q_Q$ of \eqref{eq:initialstatepauliexpansion} are then expressible as products of these random numbers $n_{\alpha}^{(i)}$, with all factors being independent of each other thanks to the local rotations being uncorrelated. Taking each pair of angles $(\theta_i,\phi_i)$ to be uniformly distributed over the sphere, one readily obtains that $\overline{n_{\alpha}}=0$ leading to $\overline{r_Q}=0$ as required by our protocol. Due to symmetry the variance $\Delta_k$ is expected to be independent of $\alpha=x,y,z$, hence we can compute it for any component. We find that
\begin{equation}
    \overline{n_z^2}= \int d\Omega  P(\theta,\phi) \cos^2(\theta) = \frac{1}{3},
\end{equation}
and thus we get a factor of a $1/3$ for each non-identity operator in a given multi-body Pauli operator $Q$. This leads to the variance $\Delta_k=1/3^k$ which in turn implies that
\begin{equation}
    \overline{r_Q^2}=\frac{1}{d} \frac{\varepsilon^{2k}}{3^k}. \label{eq:varianceaveraging}
\end{equation}

Using the result \eqref{eq:varianceaveraging} and letting $x := \epsilon^2/3$ for notational convenience, \eqref{eq:AveragedNOTOC} may be rewritten as
\begin{equation}
    F(x,t):= \overline{G(\sqrt{3x},t)} = \sum_{k=1}^N P_k(t)\, x^k, \label{eq:PGF}
\end{equation}
which is the probability-generating function (PGF) for the probability distribution $\{P_k(t)\}$ \cite{UnivariateDiscreteDistributions}. From the PGF one may extract information about the corresponding probability distribution, including the elements and moments of the probability distribution.

We point out that the uniform sampling of the continuous group SU$(2)$ is not strictly necessary, as it suffices to sample over a finite set of rotations given the correct first and second moments.
This is equivalent to constructing a unitary 2-design and sampling the local operations from it, and can be done by choosing $\mathcal{U}_\text{rot}$ such that each qubit takes on one of the values $\pm X$, $\pm Y$, and $\pm Z$ in each shot, with a total of $6^N$ unique rotation unitaries $\mathcal{U}_\text{rot}$ needed to obtain \eqref{eq:AveragedNOTOC} without approximation. This shortcut to uniform averaging also results in $\Delta_k = 1/3^k$ and $\overline{r_i}=0$, and thus \eqref{eq:PGF} is unchanged when using this method of averaging over a discrete set of rotations.

\subsection{Application of the measurement protocol}\label{subsection:epsilonprotocolnumericalresults}

\begin{figure}[t!]
    \centering
    \includegraphics[width=1\linewidth]{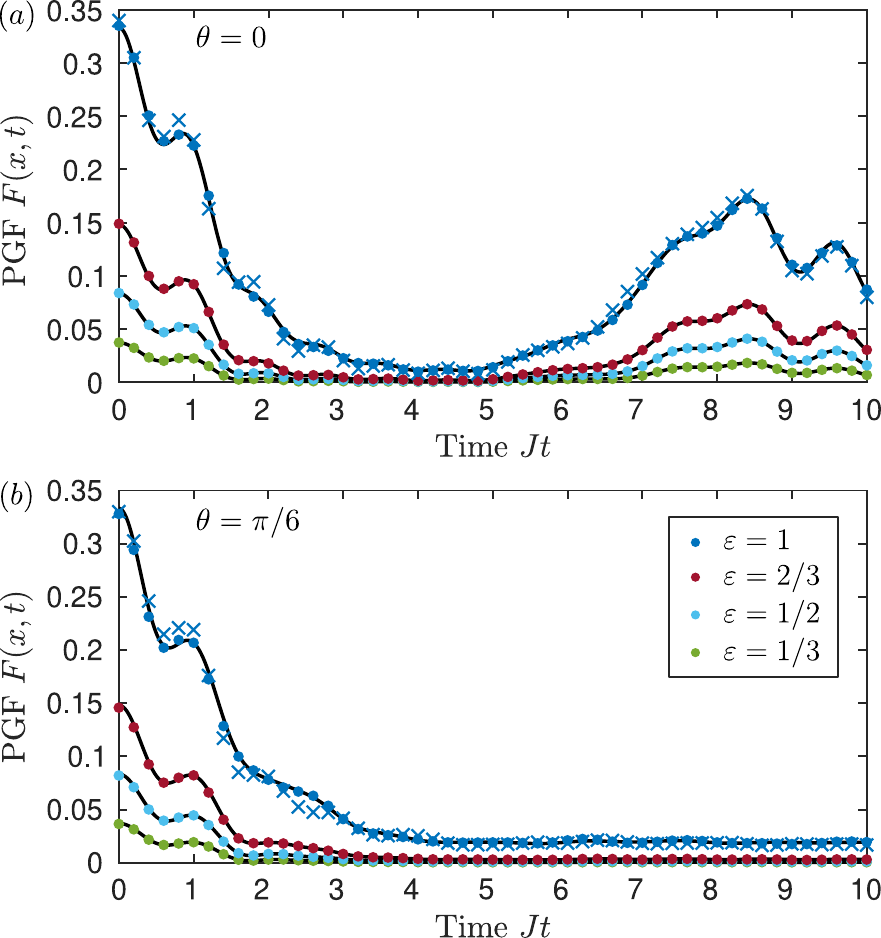}
    \caption{Exact PGF approximated using a randomized subset of rotations for $W(0) = \sigma_y^{(1)}$. Solid black lines are the exact PGF, while colored dots denote the PGF obtained using a subset of 1000 random rotations. Results are shown for several choices of the state parameter $\varepsilon$ (see legend in panel b), and for both the integrable case $\theta=0$ (a) and the chaotic case $\theta=\pi/6$ (b). For the case $\varepsilon = 1$ the blue crosses show the PGF obtained using a subset of 100 random rotations.}
    \label{fig:PGF_rotation_approx}
\end{figure}

In the remainder of this section we demonstrate via numerical simulations our measurement protocol for the case of the 1D tilted field Ising model. We present results on how to obtain the PGF approximately using a subset of random rotations, and show that the PGF may be used as an indicator of quantum information scrambling. Finally, we discuss an experimentally relevant method of extracting the elements of the probability distribution from the PGF and discuss its sensitivity to noise.

The Hamiltonian for the 1D tilted field Ising model is given by
\begin{equation}
    H_\text{Ising}(\theta) = J \sum_{i=1}^{N-1} \sigma_z^i \sigma_z^{i+1} + B \sum_{i=1}^N \left( \sigma_x^i \cos \theta + \sigma_z^i \sin\theta \right), \label{eq:IsingModel}
\end{equation}
where the operators $\sigma_\alpha^i$ are the usual Pauli operators on site $i$ with $\alpha = x,y,z$. The model describes $N$ spin-$\frac{1}{2}$ particles interacting in one dimension via nearest-neighbor interactions in the presence of an external magnetic field with both a transverse and longitudinal component which are parameterized by the angle $0 \leq \theta \leq \pi/2$. This model is a well-known platform for studies of many-body quantum chaos \cite{karthik2007entanglement,Omanakuttan2023scrambling} since it is nonintegrable for generic choices of $\theta$ and presents two integrable limits at $\theta=0$ (where the model reduces to the usual transverse field Ising model) and $\theta = \pi/2$ (where the Hamiltonian is diagonal in the $z$-basis). The scrambling properties of the dynamics generated by the Hamiltonian \eqref{eq:IsingModel} have also been studied in relation to their quantum chaos characteristics. For instance, it has been established that even the integrable limit $\theta=0$ can lead to scrambling, and in this case the mean operator size typically presents long-lived oscillations for finite system sizes. The system typically shows the highest degree of chaoticity for $\theta\simeq \pi/6$, where the operator sizes grow quickly and then equilibrate, and temporal fluctuations are suppressed \cite{Omanakuttan2023scrambling}. Throughout the following we will consider the case $J=B$ with $N=6$ qubits.

In Fig.~\ref{fig:PGF_rotation_approx}(a) and (b) the solid lines behind the colored dots show the exact PGF $F(x,t)$ for the edge site operator $W(0) = \sigma_y^{(1)} \equiv \sigma_y \otimes \mathbbm{1} \otimes \ldots \otimes \mathbbm{1}$, for several choices of the state parameter $\varepsilon$. We consider the integrable case $\theta = 0$ (a) and the chaotic case $\theta = \pi/6$ (b) of the tilted field Ising model \eqref{eq:IsingModel}. For all $t$ the two limiting cases $F(x=1,t) = 1$ and $F(x=0,t) = 0$, with $x = \varepsilon^2/3$, follow from the definition of the PGF \eqref{eq:PGF}. We observe in Fig.~\ref{fig:PGF_rotation_approx} that the primary effect of varying the state parameter $\varepsilon$ in the considered parameter range is a change of amplitude of the PGF.

The colored dots (blue crosses) in the two panels of Fig.~\ref{fig:PGF_rotation_approx} show the PGF extracted using a random subset of $1000$ ($100$) rotations out of the $6^N = 46656$ rotations needed for the exact result. Figure~\ref{fig:PGF_rotation_approx} demonstrates that we may obtain the PGF to very good accuracy using a heavily reduced number of rotations compared to the exact result, both in the integrable case $\theta = 0$ and the chaotic case $\theta = \pi/6$. This is encouraging for the experimental feasibility of implementing the present protocol.

In Ref.~\cite{Omanakuttan2023scrambling} moments of the operator size probability distribution were used as signatures of quantum information scrambling in the tilted field Ising model, with the integrable case $\theta = 0$ leading to oscillatory dynamics of the mean operator size whereas in the chaotic case $\theta = \pi/6$ the mean operator size grew and saturated only after initial oscillations. When comparing the curves for $\theta = 0$ with the corresponding curves in $\theta=\pi/6$ in Fig.~\ref{fig:PGF_rotation_approx}, we see a clear difference in behavior for $Jt \geq 4$, with $\theta = 0$ curves at later times exhibiting oscillations that are not present for $\theta = \pi/6$. In the following we thus explore whether the PGF may be used as an indicator of scrambling, similar to the mean operator size used in Ref.~\cite{Omanakuttan2023scrambling}.

\begin{figure}[t!]
    \centering
    \includegraphics[width=0.9\linewidth]{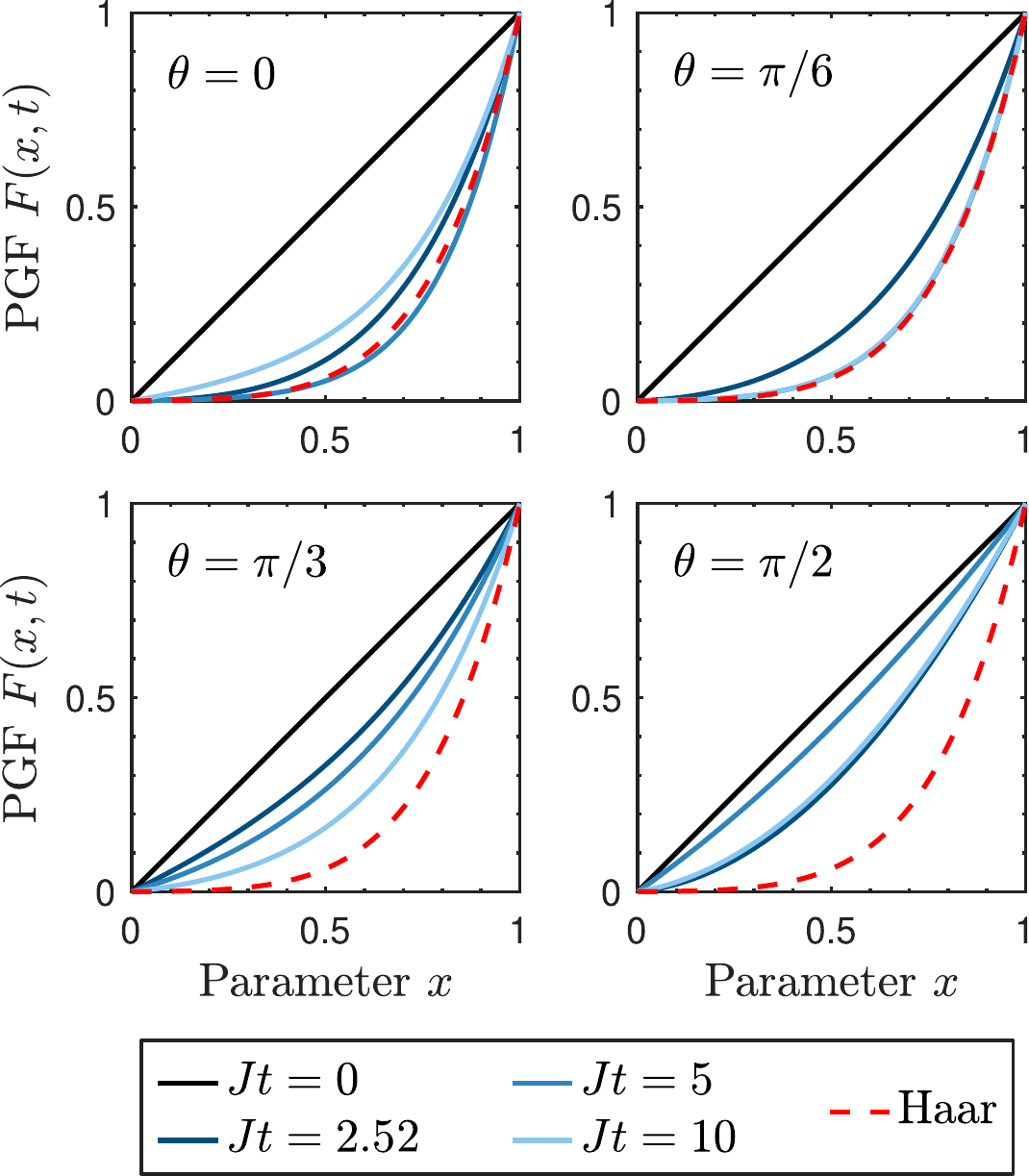}
    \caption{Exact PGF $F(x,t)$ illustrated for several times $t$ and B-field angles $\theta$ as a function of the parameter $x$. Solid lines are the PGF for $N=6$ qubits and, for comparison, the red dashed line is the PGF for a Haar random probability distribution.}
    \label{fig:PGF_shorttimes}
\end{figure}

Figure~\ref{fig:PGF_shorttimes} illustrates the PGF $F(x,t)$ for different B-field angles $\theta$ and times $t$ as a function of the PGF parameter $x$. For the initial time $t=0$ the PGF is linear in $x$ independent of the choice of $\theta$, however at finite times $t$ the behavior of the PGF is significantly different across different $\theta$s. In the chaotic case $\theta=\pi/6$ the PGF $F(x,t)$ approximately coincides with the result obtained from a Haar random state (dashed red line) for all displayed times greater than $Jt = 2.52$, indicating a fast equilibration. For $\theta = \pi/3$ the solid curves trend toward the Haar random curve for increasing time $Jt$, but have yet to equilibrate at $Jt=10$ (light blue curve). The two integrable cases $\theta = 0$ and $\theta = \pi/2$ display oscillatory behavior in the PGF $F(x,t)$ for a given $x$, similar to what was observed in \fref{fig:PGF_rotation_approx} - e.g., the dark blue $Jt=2.52$ curve is below both the black $Jt=0$ curve and the light blue $Jt = 10$ curve for all $x$ in both cases. We also do not observe an equilibration of the PGF to the result for the Haar random state.

While the PGF $F(x,t)$ primarily serves as a quantity from which one may extract information about the corresponding probability distribution, we now illustrate how the PGF itself may be used to characterize quantum information scrambling. We propose to do this by analyzing the time-average of the PGF for a fixed argument $x$,
\begin{equation}
    \overline{F}(x) = \frac{1}{t_f-t_i}\int\limits_{t_i}^{t_f} F(x,t)dt'
    \label{eq:time_avg}
\end{equation}
and the time-averaged temporal fluctuations
\begin{equation}
    \Delta F(x)^2 = \frac{1}{t_f-t_i}\int\limits_{t_i}^{t_f} \left(F(x,t')-\overline{F(x)}\right)^2 dt'.
    \label{eq:time_fluct}
\end{equation}

\begin{figure}[t!]
    \centering
    \includegraphics[width=1\linewidth]{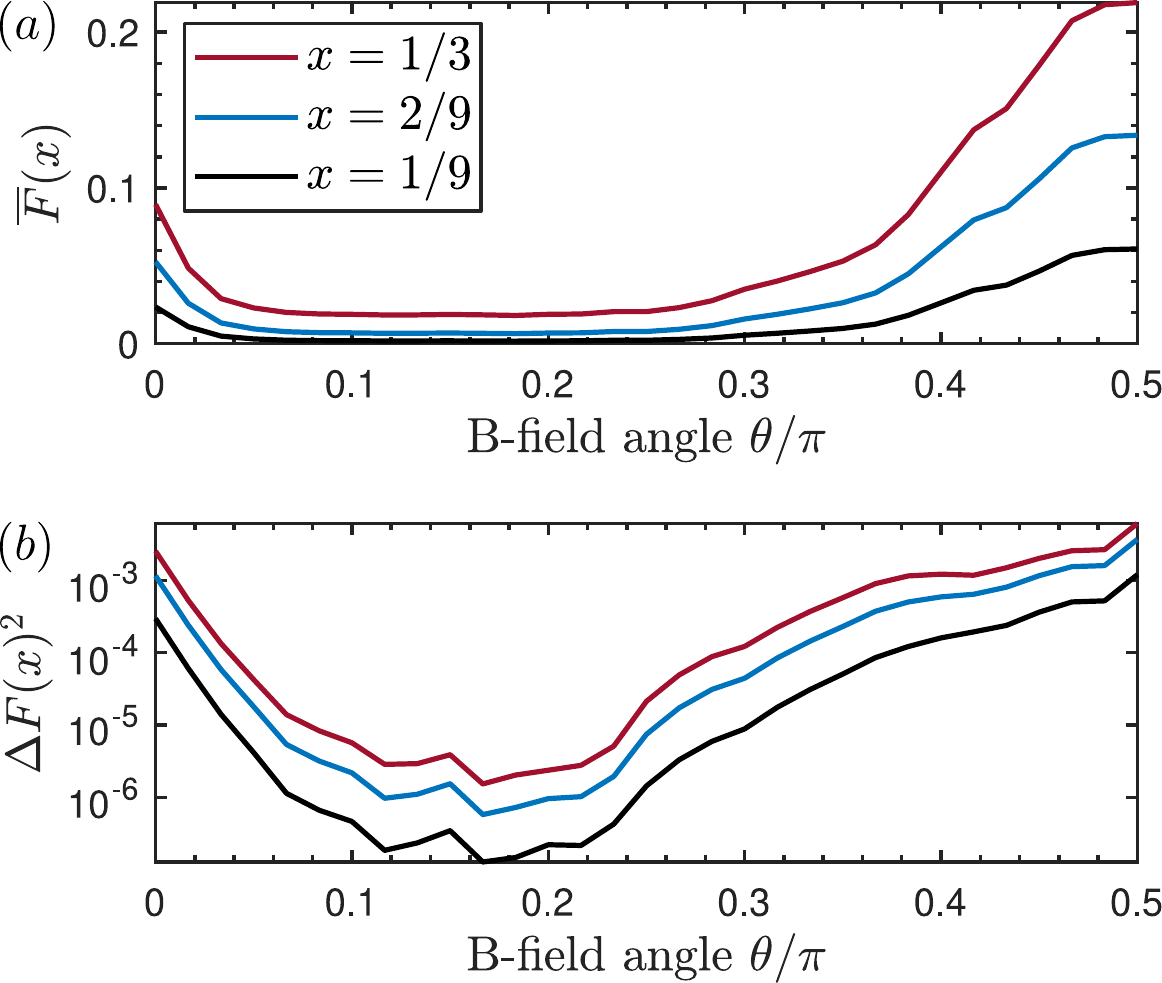}
    \caption{Time average $\overline{F}(x)$ and fluctuations $\Delta F(x)^2$ of the PGF $F(x,t)$ as a function of the B-field angle $\theta$, plotted for three values of $x$ corresponding to $\varepsilon = 1$ (red), $\varepsilon = \sqrt{2/3}$ (blue), and $\varepsilon = \sqrt{1/3}$ (black curve). The time average and fluctuations are calculated for $5 \leq Jt \leq 80$ to exclude the initial transient dynamics from the results.}
    \label{fig:PGF_scrambling}
\end{figure}

In Ref.~\cite{Omanakuttan2023scrambling} we studied analog constructions for the mean operator size (i.e., the first moment of the operator size distribution) and found that they allowed to distinguish different scrambling and quantum chaos regimes of this model. In Fig.~\ref{fig:PGF_scrambling}(a) we show the time-average $\overline{F}(x)$ as a function of the magnetic field angle $\theta$. For different accessible values of $x$ we see that that $\overline{F}(x)$ dips in the highly chaotic regime and grows near the integrable limits. This behavior originates in the fact that the chaotic case shows quick scrambling and subsequent equilibration to the Haar-random behavior, where the PGF is closer to 0, while the integrable cases show the largest typical values of the PGF, as was noted for $\theta=0$ in the discussion of Fig. \ref{fig:PGF_rotation_approx}. A similar functional form is observed for the time-averaged temporal fluctuations $\Delta F(x)^2$ which we show in Fig.~\ref{fig:PGF_scrambling}(b). This indicates that temporal fluctuations of the PGF are suppressed in the chaotic and enhanced in the integrable cases, a behavior also observed also for the mean operator size in Fig.~3 of Ref.~\cite{Omanakuttan2023scrambling}.

Our present findings thus show that it is not necessary to extract the probability distribution $\{P_k(t)\}$ from the PGF $F(x,t)$ in order to characterize the scrambling of quantum information for the considered Ising model. Rather, the time average and time fluctuations of the PGF are themselves good indicators of the presence of scrambling. As demonstrated previously in Fig.~\ref{fig:PGF_rotation_approx} the PGF can be approximated to good accuracy using a heavily reduced number of random rotations, and as the PGF is not itself sensitive to the noise unlike quantities extracted from the PGF (which we will comment on in the following Sec.~\ref{subsection:PGF_PDelement_extraction}), the PGF provides an experimentally relevant quantity for characterizing quantum information scrambling. The results of Figs.~\ref{fig:PGF_shorttimes} and \ref{fig:PGF_scrambling} further emphasize the utility of the PGF itself as a quantifier of quantum information scrambling.

\subsection{Extracting probability distribution elements and moments from the PGF}\label{subsection:PGF_PDelement_extraction}
In the absence of noise on the PGF $F(x,t)$ -- e.g., in a numerical study of the exact PGF -- the corresponding probability distribution $\{P_k(t)\}$ can be extracted from the PGF using \eqref{eq:PGF} in the following way. One first calculates the PGF $F(x,t)$ for $N$ different values of $x$, and then inverts the system of equations \eqref{eq:PGF} to obtain the elements $P_k(t)$ of the probability distribution. However, in the presence of a noisy PGF due to, e.g., a finite number of sampled rotations or experimental imperfections, we find that the procedure of inverting \eqref{eq:PGF} leads to high sensitivity to noise, which we associate with a poorly-conditioned linear system of equations.

For a more systematic approach we can use well-known properties of the PGF to recover the full probability distribution. From the definition of the PGF \eqref{eq:PGF}, one finds that \cite{UnivariateDiscreteDistributions}
\begin{equation}
    P_k(t) = \frac{1}{k!} \frac{\partial}{\partial x} F(x=0,t), \label{eq:PDelement_derivative}
\end{equation}
i.e., the elements $\{P_k(t)\}$ of the probability distribution are accessible through the evaluation of derivatives of the PGF with respect to the parameter $x$ at $x = 0$. Likewise the moments of the probability distribution may be accessed through derivatives of the PGF at $x=1$:
\begin{equation}
    \mathbb{E}[X^k(t)] = \left(x \frac{\partial}{\partial x}\right)^k F(x=1,t). \label{eq:moments_derivative}
\end{equation}
In the following we will focus on the extraction of the elements of the probability distribution via \eqref{eq:PDelement_derivative}. We leave the discussion of the moment extraction for later in this section.

The $n$th derivative of the PGF $F(x,t)$ at $x=0$ may be implemented for both numerical studies and in experiments using a forward-only finite difference method on the form \cite{NumericalMethodsForOrdinaryDifferentialEquations}
\begin{align}
    F^{(n)}(x=0,t) =& (\Delta x)^{-n} \sum_{m=0}^{n+a-1} c_m^{(n,a)} F(m\, \Delta x, t) \nonumber\\
    &+ \mathcal{O}(\Delta x^{a}), \label{eq:forwardonlyfinitediff}
\end{align}
where $a$ is the accuracy of the finite difference method. We leave the details of our implementation of the forward-only finite difference method to appendix~\ref{appendix:PDelements_finitedifference}. In practice, applying this method implies preparing several states of the form \eqref{eq:polarizedqubitstate} with various levels of purity (controlled by the parameter $\varepsilon$). To explore the experimental feasibility of extracting the probability distribution elements from the PGF, in the following we characterize the sensitivity of the method to noise. The noise may have origin in an approximative PGF due to choosing a reduced subset of rotations, as discussed in Sec.~\ref{subsection:epsilonprotocolnumericalresults}, or stem from experimental imperfections. We simulate the effect of the noise in the following way. First the exact PGF $F(x,t)$ is calculated, on top of which we add noise. The noise is assumed to be Gaussian and multiplicative in nature \footnote{If one instead assumes that the noise is additive, i.e. $F_\eta(x,t) = F(x,t) + \delta(\eta)$, one finds an increased sensitivity to noise compared to the multiplicative case due to the value of the PGF being close to zero whenever the parameter $x$ is also close to zero, hence the impact of additive noise on the PGF is higher than multiplicative.}, hence we write the noisy PGF as
\begin{equation}
    F^\eta(x,t) = \left[ 1 + \delta(\eta) \right] F(x,t), \label{eq:multiplicative_noise}
\end{equation}
where $\delta(\eta) \sim \text{N}(0,\eta)$ is a normal-distributed number with vanishing mean and variance $\eta^2$. From the noisy PGF we then use \eqref{eq:forwardonlyfinitediff} to extract the elements $\{P_k^\eta(t)\}$ of the probability distribution in the presence of noise.

\begin{figure}
    \centering
    \includegraphics[width=1\linewidth]{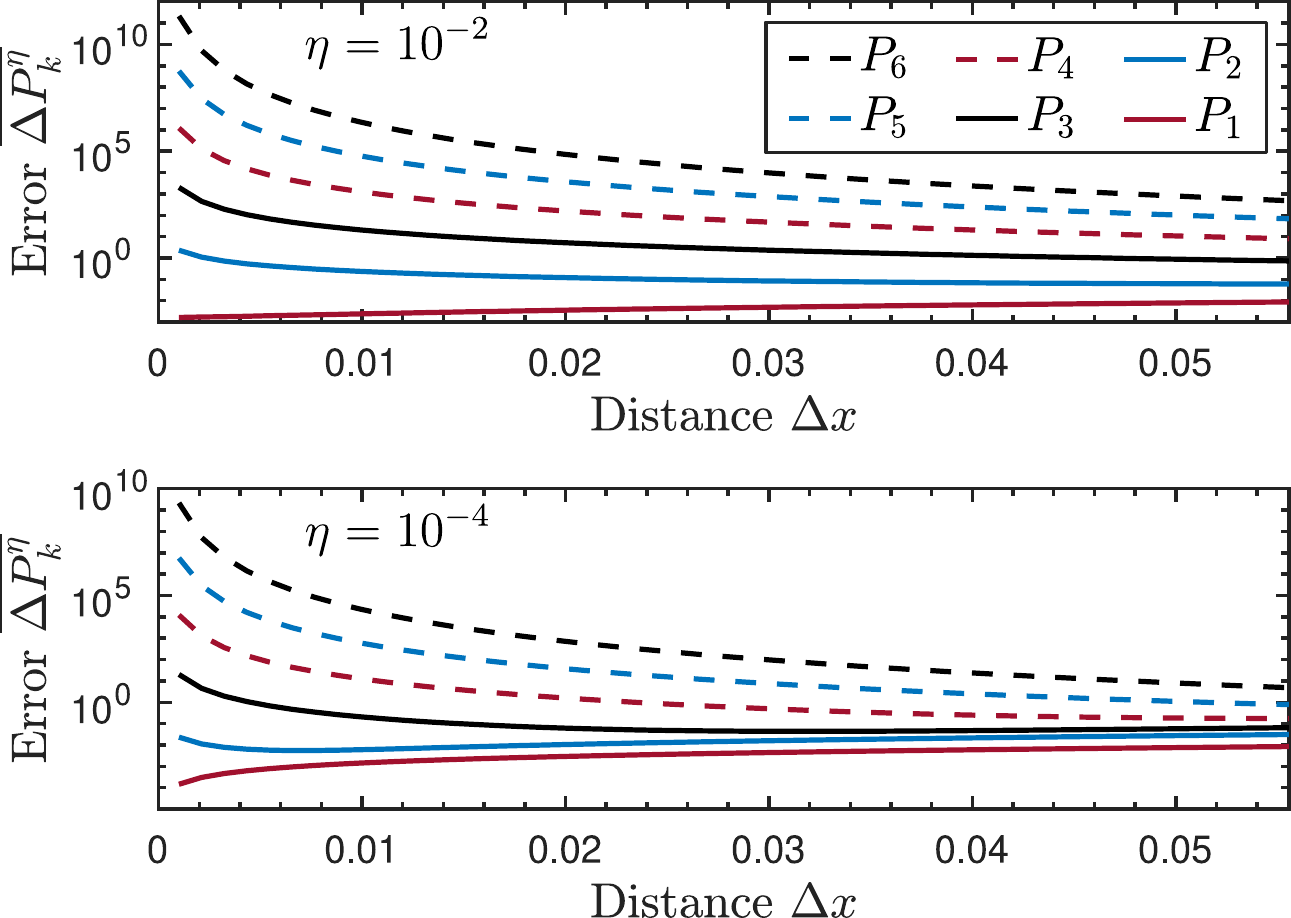}
    \caption{Time averaged error $\overline{\Delta P_k^\eta}$ [see \eqref{eq:TimeAveragedError}] for the elements of the probability distribution $\{P_k\}$, averaged over $0\leq Jt \leq 10$. We have used multiplicative noise with strength $\eta = 10^{-2}$ (upper panel) and $\eta = 10^{-4}$ (lower panel), added to the PGF $F(x,t)$ before using finite difference derivatives with step size $\Delta x$ and accuracy $a=2$ to extract the noisy probability distribution elements $\{P_k^\eta(t)\}$. In both panels the curves from top to bottom at $\Delta x = 0.02$ are the time averaged errors $\overline{\Delta P_k^\eta}$ for $P_6$ (black dashed), $P_5$ (blue dashed), $P_4$ (red dashed), $P_3$ (black solid), $P_2$ (blue solid), and $P_1$ (red solid), averaged over 100 noise realizations.}
    \label{fig:derivative_individual}
\end{figure}

We now consider the $N=6$ transverse field Ising model with $J=B$ and $\theta = 0$. In Fig.~\ref{fig:derivative_individual} we show for the operator $W(0)=\sigma_y^{(1)}$ the time averaged error of the extracted probability distribution elements
\begin{equation}
    \overline{\Delta P_k^\eta} = \frac{1}{T} \int_0^T \mathrm{d}t \vert P_k^\eta(t) - P_k^0(t) \vert, \label{eq:TimeAveragedError}
\end{equation}
as a function of the finite difference step size $\Delta x$. The error is calculated by comparing the elements $P_k^\eta(t)$ extracted from the noisy PGF with the exact elements $P_k^0(t)$, and has been averaged over 100 realizations of the Gaussian distributed noise $\delta(\eta)$. An accuracy of $a=1$ was used for the finite difference method; see details in appendix~\ref{appendix:PDelements_finitedifference}.

When using finite-difference methods, there are two major sources of error: truncation errors associated with the accuracy $a$ of the method, and rounding errors due to uncertainty on the input function (in the present case the rounding error is due to the noise added to the PGF) \cite{NumericalMethodsForOrdinaryDifferentialEquations}. The former error prefers $\Delta x$ as small as possible, while the latter prefers larger $\Delta x$. Hence, in choosing the step size $\Delta x$, one has to balance the contributions from the two errors. We observe for the case $\eta = 10^{-2}$ in Fig.~\ref{fig:derivative_individual} that there exists step sizes $\Delta x$ for which the first two elements $P_1$ and $P_2$ can be extracted with reasonable errors -- for $P_1$ the time-averaged error is much smaller than unity for all choices of distance $\Delta x$. For higher $k$ the errors exceed $10^{-1}$ for all choices of $\Delta x$, and is dominated by rounding errors~\cite{NumericalMethodsForOrdinaryDifferentialEquations}. We note here that the step size $\Delta x$ is upper bounded by $(1/3)/(n+a-1)$, as we need to sample $(n+a-1)$ points between $x=0$ and $x = 1/3$.

By decreasing the noise amplitude by two orders of magnitude to $\eta = 10^{-4}$, we observe in the lower panel of Fig.~\ref{fig:derivative_individual} that choosing $\Delta x \geq 0.03$ yields time averaged errors smaller than $10^{-1}$ for the $k\leq 3$ elements, while the $k \geq 4$ elements of the probability distribution still carry too large errors to be usable for the analysis of scrambling. We thus see that the above derivative-based method \eqref{eq:PDelement_derivative} of extracting the elements of the probability distribution becomes increasingly sensitive to noise with the degree of the derivative one needs to evaluate, and even for highly precise measurements of the PGF $F(x,t)$ the method \eqref{eq:PDelement_derivative} of extracting the probability distribution from the PGF is too sensitive to noise to allow us access to $P_k$ for $k \geq 4$.

As a consequence of the high sensitivity to noise -- whether the noise originates in experimental imperfections or an approximate PGF due to a finite number of sampled rotations -- it does not seem experimentally feasible to extract the probability distribution elements from the PGF. Instead, one should use a protocol dedicated to the extraction of the probability distribution elements. To this end we present in Sec.~\ref{section:nonepsilonprotocol} an experimentally-relevant protocol that is able to access the elements of the probability distribution without using the PGF.

Finally, we comment on the extraction of the moments of the probability distribution using \eqref{eq:moments_derivative}. Accessing the moments of the probability distribution using the previously described finite difference method requires one to create states with $x > 1/3$ to approximate the derivative. However, with the constraint of $\vert\varepsilon\vert \leq 1$, it is impossible to access $x > 1/3$ through the procedure outlined in this section. If one could engineer an ensemble of correlated initial states, the variance $\Delta_k$ could possibly be manipulated sufficiently to allow access to values of $x > 1/3$. However, in the present work we do not investigate this further and leave it for possible future extensions of this work.

\section{Method B: accessing elements of the probability distribution for small operator sizes}\label{section:nonepsilonprotocol}

\begin{figure}
    \centering
    \includegraphics[width=\linewidth]{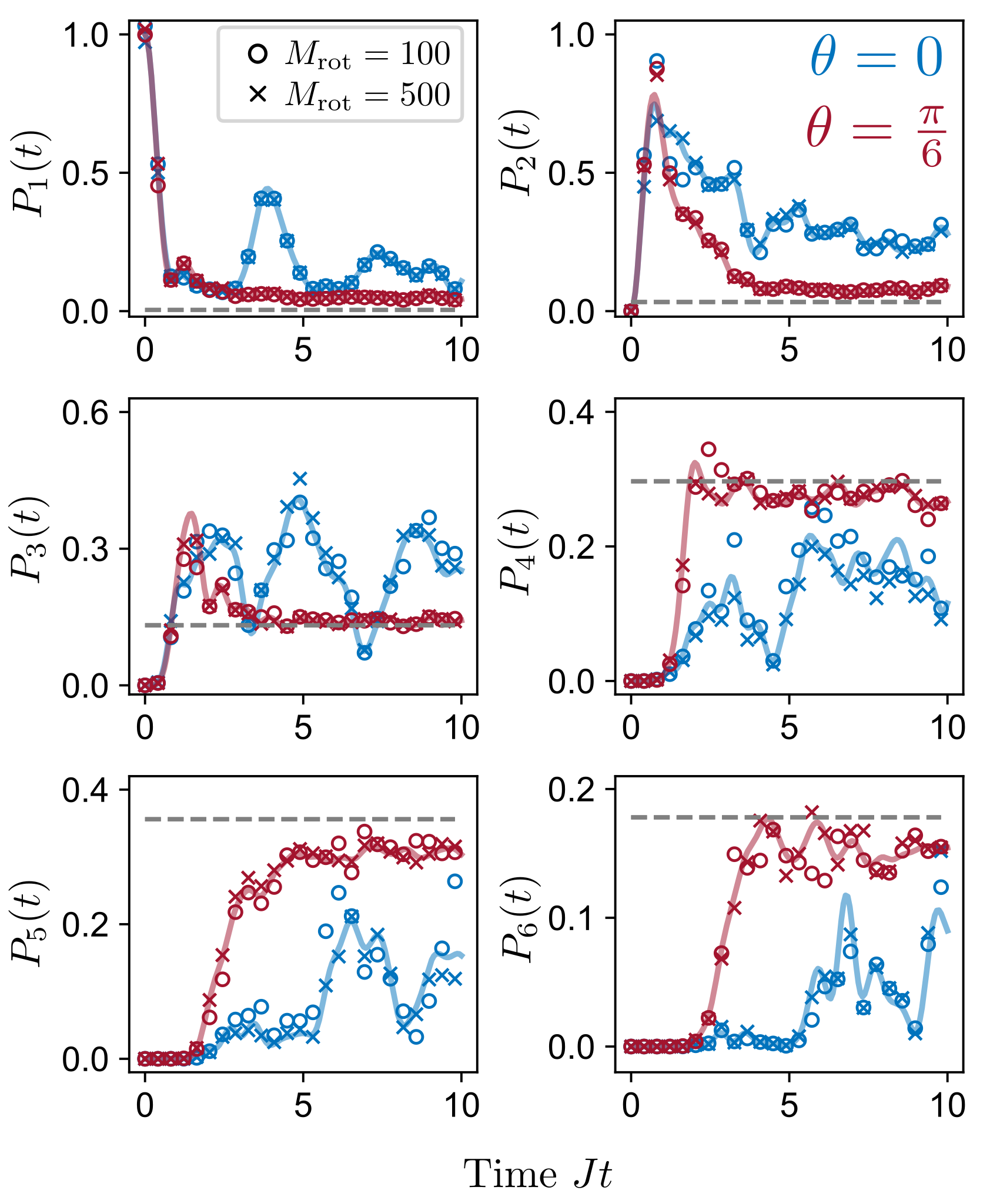}
    \caption{Application of Method B to obtaining the operator size distribution $\{P_k(t)\}$ for the case of the Ising model of \eqref{eq:IsingModel} with $N=6$ and $J/B=1$. Each panel shows a different value of $k=1,\ldots,6$. Full lines correspond exact numerical results obtained by solving the Heisenberg evolution of the initial operator, here chosen to be $W(0)=\sigma_x^{(2)}$. Symbols correspond to numerical simulations of Method B (Sect. \ref{section:nonepsilonprotocol}) with different choices of the number of sampled rotations $M_{\text{rot}}=100$ (circles) and $M_{\text{rot}}=500$ (crosses). Different colors denote different regimes of the model: integrable case $\theta=0$ (blue) and chaotic case ($\theta=\pi/6$) (red).}
    \label{fig:noneps_figure}
\end{figure}

In the previous Sec.~\ref{section:epsilonprotocol} we presented a measurement protocol that can access the operator size probability distribution $\{P_k(t)\}$ by measuring the probability-generating function (PGF) $F(x,t)$ in \eqref{eq:PGF}.
Here we return to the general result in \eqref{eq:NOTOC} and develop an alternate NOTOC measurement protocol to access the individual probabilities.
This is achieved by employing a different choice of initial state that replaces the one in \eqref{eq:polarizedqubitstate}. 
The protocol will provide a direct way of obtaining the probabilities and bypasses the sensitivity issues encountered when trying to invert the PGF. 

In order to directly access $P_k$ for a given $1\leq k \leq N$, we choose a subset $\mathcal{M}_k$ of $k$ particles in the system and prepare them the classically correlated state
\begin{equation}
    \rho_Z^{(k)} = \frac{1}{2^k} \left( \mathbb{I}_k +  \bigotimes\limits_{i=1}^{k} Z_i\right),
    \label{eq:noneps_rho0}
\end{equation}
while the other $N-k$ particles are left in a maximally mixed state. The resulting state is clearly unentangled as it can be produced as a statistical mixture of product states in the computational basis. Crucially, it has the desired property of being written solely in terms of weight-$k$ operators, as discussed in Sec.~\ref{section:protocol}. Thus we may proceed in a similar fashion to method~A: we apply a round of uniformly-chosen random rotations on each of the particles in $\mathcal{M}_k$. The resulting full state takes the form
\begin{equation}
    \rho_0 = \frac{1}{d}\left(\mathbbm{1} + \sum\limits_{Q\in C_k(\mathcal{M}_k)} q_Q Q\right),
\end{equation}
where the modified set $C_k(\mathcal{M}_k)$ is composed of all $k$-body Pauli operators acting on the particles in $\mathcal{M}_k$. The application of random, local rotations ensures that the coefficients $q_Q$ are independent, identically distributed random variables with mean $\overline{q_Q}=0$ and variance $\overline{q_Q^2}=1/3^k$. It follows from \eqref{eq:NOTOCp} that
\begin{equation}
    \overline{G(t)}_{\mathcal{M}_k} = \frac{1}{3^k} \sum\limits_{Q\in C_k(\mathcal{M}_k)} \left\vert f\left[Q;W(t)\right]\right\vert^2,
\end{equation}
and thus we find
\begin{equation}
    P_k(t) = 3^k\sum\limits_{\mathcal{M}_k} \overline{G(t)}_{\mathcal{M}_k}.
    \label{eq:Pk_noneps}
\end{equation}
In order to obtain $P_k(t)$ exactly one needs to obtain $G(t)$ for all possible subsets of $\mathcal{M}_k$ of $k$ particles, thus requiring $N\choose{k}$ repetitions of the protocol. While this is in general inefficient, it can be feasible as long as $k$ and $N$ are not too large. In practice, the number of initial states can be reduced in at least two ways. One option is to approximate the sum in \eqref{eq:Pk_noneps} by randomly sampling a reduced number of sets $\mathcal{M}_k$. Alternatively one can exploit symmetries of the system. For instance, the Hamiltonian in \eqref{eq:IsingModel} has a reflection symmetry with respect to the middle of the chain which can be easily leveraged to reduce the state count by a factor of $2$. If the system has a full translational symmetry, then the state count can be reduced by a factor of $N$. The procedure to achieve this is described in the Appendix \ref{app:symmetry}.

In the following we present numerical results that demonstrate this method's ability to reconstruct $P_k(t)$ for various instances of the tilted field Ising model introduced in Sec.~\ref{subsection:epsilonprotocolnumericalresults} with $N=6$ particles. We choose the initial operator to be $W(0)=\sigma_x^{(2)}$. Results are obtained by using all possible choices of $\mathcal{M}_k$ for each $k$, and by sampling over $M_{\mathrm{rot}}$ realizations of randomized rotations. In \fref{fig:noneps_figure} it can be seen that the dynamics of each size probability $P_k(t)$ is faithfully reproduced by the present method, in some cases using as little as $M_{\mathrm{rot}}=100$ rotations. For larger operator sizes $k\geq 4$ the effects of finite sampling are more pronounced, however this is to be expected as the typical probabilities are also smaller. The cases displayed in the figure correspond to $\theta=0$ (integrable transverse field model, blue curves and symbols) and $\theta=\pi/6$ (chaotic model, red curves and symbols), and it can be readily seen that the protocol accesses the typical features expected in both cases, namely a fast spread of the operators (as seen in the decay of $P_1(t)$ at short times), a subsequent oscillatory behavior for $\theta=0$, and equilibration for $\theta=\pi/6$.

We note that the method presented in this section will not be feasible to obtain the full probability distribution for large system sizes $N\gg 1$, since at some point an exponential number of the subsets $\mathcal{M}_k$'s would be required to exactly recover them from \eqref{eq:Pk_noneps}. However, this method can be used to verify whether a system a system has scrambled to $k-$body operators as a long as $k\sim O(1)$. An alternative benchmark is to compare the values of the obtained $P_k$'s with the ones corresponding to Haar-random evolution, i.e.
\begin{equation}
    P_k^{\text{Haar}}={N\choose k} \frac{3^k}{d^2-1}
\end{equation}
where $d=2^N$ (see Refs.~\cite{qi2019,Omanakuttan2023scrambling} for additional details). We show these values as gray dashed lines in each of the plots of \fref{fig:noneps_figure}, where we observe that the observable tends to equilibrate to these values in the chaotic regime ($\theta=\pi/6$) of the model.

\section{Comparison \& discussion}\label{section:discussion}

\subsection{Extensions to other systems}
In this section we present an extension of the NOTOC measurement protocol introduced in Sec. \ref{section:protocol} to collective spin systems, where every spin interacts with every other spin in the system. 
This class of systems preserves $J^{2}=J_{x}^{2}+J_{y}^{2}+J_{z}^{2}$ where $J_{i}=\sum_{k}\sigma_{i}^{(k)}$ is the collective spin operator. Some well known Hamiltonians of this type include p-spin models ($p=2$ case is referred to as the Lipkin-Meshkov-Glick model) and the quantum kicked top model \cite{Meshkov1965,Ribeiro2007,Haake1987}. The dynamics of these models is often studied in the symmetric subspace, the subspace of the whole Hilbert space associated with $J=N/2$ quantum number where $N$ is the number of spins. The dimension of this subspace increases linearly with the number of spins in the system, $d_{ss}=2J+1=N+1$.

It is natural to study scrambling in this type of system by decomposing the time-evolved operators in a basis associated with the polynomials of collective spin operators, referred to as the spherical tensor operator basis \cite{Klimov2008,Omanakuttan2023scrambling}. A spherical tensor operator $T^{(k)}_{q}$ is an operator that transforms under rotations in the same manner as spherical harmonics \cite{Ballentine2014}
\begin{align}
    \mathcal{D}(\alpha,\beta,\gamma)T^{(k)}_{q}\mathcal{D}^{\dagger}(\alpha,\beta,\gamma)&=\sum_{q'=-k}^{k} \mathcal{D}_{q'q}^{(k)}(\alpha,\beta,\gamma)T^{(k)}_{q'},
\end{align}
where $\mathcal{D}(\alpha,\beta,\gamma)= e^{-iJ_{z}\alpha}e^{-iJ_{y}\beta}e^{-iJ_{z}\gamma}$ and therefore $\mathcal{D}_{q'q}^{(k)}(\alpha,\beta,\gamma)\equiv \langle k,q'|\mathcal{D}(\alpha,\beta,\gamma)|k,q\rangle$. The explicit form of the spherical tensor operators is given by \cite{Klimov2008}
\begin{align}
T_{q}^{(k)}(J)&= \sqrt{\frac{2k+1}{2J+1}} \sum_{m,m'=-J}^{J} C_{J\; m;k\; q }^{J\; m'} |J,m'\rangle \langle J,m| ,
\end{align}
where  $C_{J\; m;k\; q }^{J\; m'}=\langle J,m'|J,m;k,q\rangle$ is a Clebsch–Gordan coefficient, and $q=\{-k,-k+1,...,k\}$ for a given rank of the tensor operator $k=\{0,1,...,N\}$. These operators form an orthonormal basis $\text{Tr} \left[(T^{(k')}_{q'})^{\dagger}T^{(k)}_{q}\right]=\delta_{k,k'}\delta_{q,q'}$ that spans the Hilbert space associated with the symmetric subspace. Note that a $k$th rank tensor operator is simply a $k$th order polynomial of collective spin operators, hence the basis consists of polynomials of collective spin operators ranging from order $0$ to order $N$. The rank of these operators can be used to construct the operator size distribution for this class of systems, similar to the operator size (Hamming weight) used for the Pauli basis. It is then natural to analyze the dynamics of the system by considering an operator of a particular rank $k$ at the initial time, and characterizing the scrambling dynamics of the system by considering for each tensor rank $k$ the time evolution of the element $P_k(t)$ of the operator size distribution.

In the following we describe the details of a NOTOC measurement protocol for collective spin systems that can access the elements $P_k(t)$ of the operator size distribution for small operator sizes. As in Sec.~\ref{section:protocol} we require the initial state to be prepared from an ensemble that satisfies $\overline{r_{i}}=0$ and $\overline{r_{i}^{2}}=\Delta_{k}$ in \eqref{eq:initialstateexpansion}. The nonzero coefficients $r_i$ are here associated with a particular rank $k$ in the expansion of the density operator in the spherical tensor operator basis $\{T_{q}^{(k)}\}$.
To obtain states that satisfy these properties for the expansion coefficients, we start with a mixed state given by
\begin{align}
\label{eq:fully_connected_state}
\widetilde{\rho}_{0}&=\frac{1}{d_{ss}}\biggl(\mathbbm{1}+\frac{1}{e_{0}}T^{(k)}_{0}\biggr),
\end{align}
where $k>0$. $e_{0}$ is the absolute value of the smallest eigenvalue of $T_{0}^{(k)}$, which is used to ensure the semidefiniteness of the density operator. Note that $T^{(k)}_{0}$ is a $k$th order polynomial in $J_{z}$. For instance, $T_{0}^{(1)}=c_{0}^{(1)} J_{z}$ and $T_{0}^{(2)}=c_{2}^{(0)} (3J_{z}^{2}-J(J+1))$ where $c_{0}^{(1)}$ and $c_{0}^{(2)}$ are normalization coefficients.  For $k=1$ the state $\widetilde{\rho}_0$ is a thermal equilibrium state of a system placed in an external magnetic field at high temperatures, $\rho =\frac{1}{d}(1+\epsilon J_{z})$ where $\epsilon \ll 1$ \cite{Wei2018}. Higher-order states could potentially be prepared as equilibrium states of a system with Hamiltonian consisting of higher-order $J_{z}$ terms.

Applying global rotations of the form $D(\phi,\theta)=\mathcal{D}(\alpha=\phi,\beta=\theta,\gamma=0)= e^{-iJ_{z}\phi}e^{-iJ_{y}\theta}$ on the state $\widetilde{\rho}_0$ leads to
\begin{align}
\begin{split}
    \rho_{0}&=D(\phi,\theta)\; \tilde{\rho_{0}}\; D^{\dagger}(\phi,\theta) \\
&= \frac{1}{d_{ss}} \biggl(\mathbbm{1}+\frac{1}{e_{0}}\sum_{q'=-k}^{k} \langle k, q'|D(\phi,\theta)|k,0\rangle \; T_{q'}^{(k)}\biggr)\\
& \equiv \frac{1}{d_{ss}}\mathbbm{1}+\sum_{q'=-k}^{k} r_{k,q'} \; T_{q'}^{(k)}
\end{split}
\end{align}
where $r_{k,q'}=(e_{0}d_{ss})^{-1}\langle k, q'|D(\phi,\theta)|k,0\rangle$, and the angles $\theta$ and $\phi$ are sampled randomly from a uniform distribution on the surface of a sphere. This state has zero mean and nonzero variance \cite{Varshalovich1988} as expected,
\begin{align}
    \overline{r_{k,q'}}&=\int d\Omega \; r_{k,q'}(\phi,\theta)=\frac{1}{e_{0}d_{ss}}\overline{\langle k, q'|D(\phi,\theta)|k,0\rangle}= 0, \\
    \begin{split}
    \overline{r_{k,q'}^{2}}&=\int d\Omega \; r_{k,q'}^{2}(\phi,\theta)=\frac{1}{e_{0}^{2}d_{ss}^{2}} \overline{|\langle k, q'|D(\phi,\theta)|k,0\rangle|^{2}}\\
    &= \frac{1}{e_{0}^{2}d_{ss}^{2}} \frac{4\pi}{2k+1},
    \end{split} \label{eq:collective_rk}
\end{align}
where $d\Omega =\sin{\theta} d\theta d\phi$. Equation~(\ref{eq:collective_rk}) is analogous to \eqref{eq:varianceaveraging} in that the expression for the variance depends only 
on a coarse-grained property of the operator basis element, namely, the operator size for the Pauli basis and the operator rank for the spherical tensor basis.
Hence we have 
\begin{align}
    P_{k}(t)=\frac{1}{\Vert W \Vert_2^2\;\overline{r_{k,q'}^{2}}}  \overline{\vert \braket{W(t)}\vert^2}
\label{eq:noneps_collective}
\end{align}
where $W$ is the operator of interest. This result shows that method B of Sec.~\ref{section:nonepsilonprotocol} can be naturally adapted to collective spin systems. In general, the procedure shown in this section illustrates how to choose a combination of initial states and randomized operations tailored to the choice of operator basis such that the average NOTOC connects to operator size distributions. Notice that $N+1$ repetitions of the above protocol for different $k$ will provide us probabilities associated with all operator sizes. However, the state in \eqref{eq:fully_connected_state} is not easily accessible for higher values of $k$, so this protocol might only be suitable for small system sizes.

\subsection{Relation to previous proposals}
The toolbox presented in Sec.~\ref{section:protocol} presents some noteworthy connections with previous works which have studied how to diagnose complex many-body dynamics in different settings. For instance, Qi \etal \cite{qi2019} proposed a method to probe the growth of an operator $O$ in quantum quench experiments using pure product states of qudits (of local dimension $d_L$) followed by random local operations. The authors showed that the variance $\delta O(t)^2$ of the expectation value $\langle O(t)\rangle$ over the random realizations yields
\begin{equation}
    \delta O(t)^2 = \frac{\text{Tr}(O^2)}{d_L^N}F\left(x=\frac{1}{d_L+1},t\right).
\end{equation}
where $F(x,t)$ is the probability-generating function associated with the operator size distribution of $O(t)$. For qubits, $d_L=2$ and so the method probes the PGF at $x=1/3$. This is exactly the case for the NOTOC when choosing states of the form proposed in method A (Sec.~\ref{section:epsilonprotocol}) using pure states ($\varepsilon =1$), see \eqref{eq:PGF}. Therefore, our proposed method A recovers the protocol of Ref.~\cite{qi2019} for the case of pure states and generalizes it by showing that the PGF can be sampled in a continuum of values $x\leq 1/3$ by using mixed states of qubits. It remains to be studied whether this generalization carries over to the case of qudits with $d_L>2$. Additionally, these methods do not allow to access $x>1/3$ directly and it is unclear whether the toolbox of Sec.~\ref{section:protocol} provides a way around this by using a clever choice of initial states. Regarding this aspect, we point out that a recent work proposes an alternative method to access the PGF $F(x,t)$ (in principle for any $x$) using a single-particle mixed states (similar in form to \eqref{eq:noneps_rho0} when $k=1$) and resorting to forward $U(t)$ \textit{and} backward $U^\dagger(t)$ evolution. Importantly, the analysis we presented in Sec.~\ref{subsection:PGF_PDelement_extraction} concerning the large sensitivity to noise of the process of obtaining the operator distribution $\{P_k(t)\}$ from its PGF $F(x,t)$ applies to all methods that aim at obtaining the PGF. Our findings indicate that obtaining the distribution from the PGF might be unfeasible in experiments, but also show that properties of the PGF itself could be used a probe for scrambling directly. More detailed work should be carried out to explore this further.

The idea of using mixed states to probe properties of operator evolution was also used recently by Peng \etal in Ref.~\cite{arXiv.2209.09322}, where the goal was to measure single-site two-time correlation functions on the form $\sum_i \text{Tr}[\sigma_z^{(i)}(0)\, \sigma_z^{(i)}(t)]$ in an NMR experiment. This measurement was carried out by first preparing the weakly polarized initial state $\rho_0 \propto (\mathbbm{1} + \varepsilon \sum_i \sigma_z^{(i)})$, then allowing this initial state to become locally randomized by the effect of on-site disorder, and ultimately measuring a tunable observable using inductive measurements, rotations, and on-site disorder. An analogous method for measuring two-site two-time correlation functions was also proposed by the authors. The measurement protocol proposed in Sec.~\ref{section:epsilonprotocol} of the present work shares significant overlap in methodology with that of Ref.~\cite{arXiv.2209.09322}, however the goals of the two measurement protocols differ and thus the prepared random mixed states and ultimate measurements are also different. For the purpose of extracting the operator size distribution, we note that by using the two-time correlation function studied in Ref.~\cite{arXiv.2209.09322} one will have to extend the method of Ref.~\cite{arXiv.2209.09322} to all $m$-site correlation functions, with $m\leq N$. This will likely not be feasible due to the non-local nature of the operators to be measured, as well as the issue of the exponentially growing number of operators one needs to measure which was discussed in Sec.~\ref{section:nonepsilonprotocol} for our proposed measurement protocol.

Finally, we comment on the connection between our proposal and that of Vermersch \etal in \cite{PhysRevX.9.021061}, where the authors propose a way to measure OTOCs without using time-reversal operations or auxiliary systems. Instead, their proposal is based on performing randomized unitaries on a set of initial states and extracting the OTOCs from the statistical correlations between the measurement results. In principle, this method allows one to reconstruct the operator size distribution if one repeats the procedure for (exponentially many) choices of the operator $R$ in \eqref{eq:standardOTOC}. This can be achieved by using averages of OTOCs to obtain the moments of the $\{P_k(t)\}$ distribution, as outlined in Sect. VI of \cite{Omanakuttan2023scrambling}. In contrast, our method can be seen as a way of using similar tools (i.e. preparation of product states, randomized local operations, and local measurements) to probe the operator size distribution \textit{directly}, thus bypassing the calculation of OTOCs.

\subsection{Relation to fidelity OTOCs}
Finally, we discuss the physical interpretation of the different tools and quantities used to study quantum information scrambling.
The first aspect is related to the fidelity OTOCs, introduced in Sec.~\ref{section:protocol}, which are a class of correlation functions obtained from the usual OTOCs of the form in \eqref{eq:standardOTOC} by choosing the early-time operator $R$ to be the projector onto the initial state $R=\rho_0$.
The use of fidelity OTOCs attracted widespread attention in the community because they are \textit{technically} an OTOC but can be measured as a single expectation value of a (often) local operator $W$ when the initial state is pure, $\rho_0=\ket{\psi_0}\bra{\psi_0}$ \cite{PhysRevA.106.042429}. 
Fidelity OTOCs have interesting connections to quantities like the quantum Fisher information \cite{NatCommun.10.1581} and the Loschmidt echo \cite{gorin2006dynamics}.

The fact that typically $\rho_0$ is a non-local operator makes the fidelity OTOC relinquish the usual interpretation of OTOCs as measures of information scrambling. 
In particular, the relation between OTOCs involving Pauli operators and moments of the operator size distributions \cite{zhuang2019scrambling,Omanakuttan2023scrambling} does not apply to fidelity OTOCs.
However, our present work shows that fidelity OTOCs are indeed connected fundamentally to operator size distributions if one takes the initial pure state $\ket{\psi_0}$ to be a product state and then considers the average fidelity OTOC over many realizations of local random rotations on the initial state.

\section{Conclusion}\label{section:Outlook}
In this article we have proposed a measurement protocol for probing quantum information scrambling by measuring operator size distributions. Our measurement protocol requires the preparation of separable mixed states followed by local operations and final-time measurements of local operators, and circumvents the typical use of out-of-time-ordered correlation functions (OTOCs) to probe scrambling properties\cite{PhysRevA.94.040302,PhysRevA.99.051803}. We have demonstrated that the choice of initial separable mixed states in our measurement protocol provides multiple ways to access the operator size distribution, and we comment on the experimental feasibility for two particular methods, one based on first extracting the probability-generating function for the operator size distribution, and the other focused on obtaining the elements of the operator size distribution directly.

The application of our measurement protocol is illustrated in detail for the 1D tilted field Ising model, a well-known platform for studying many-body quantum chaos \cite{prosen2002general,PhysRevA.71.062324,karthik2007entanglement}. We numerically demonstrate the characterization of quantum information scrambling for this model using our proposed measurement protocol. We have related our proposed protocol with well-established methods of characterizing scrambling, including those based on OTOCs, and were able to establish connections between the well-established fidelity OTOC and operator size distributions using our results presented in this article. Finally, we have exemplified the extension of our measurement protocol to other types of quantum systems by considering our protocol for the case of collective spin systems. The collective spin case emphasizes further the role that state preparation plays in our measurement protocol.

In the discussion of our proposed measurement protocol's connection to the probability-generating function, we found that the extraction of moments of the operator size distribution was not possible due to constraints on the prepared initial states that prevent us from accessing $x>1/3$ in the probability-generating function $F(x)$. From preliminary numerical results we expect that the extraction of the first and second moments of the operator size distribution would be both resilient to noise and reasonable to implement in experiment if one could obtain values of $x > 1/3$, hence finding a way to prepare initial states allowing the extraction of moments of the operator size distribution would be an obvious extension of the present work. This potential extension should be compared to the proposal in Ref.~\cite{schuster2022} where the probability-generating function can be accessed for any $x$, but at the expense of requiring the implementation of time-reversal of the many-body evolution.

In Sec.~\ref{section:discussion}A we presented the extension of our NOTOC measurement protocol to collective spin systems. The extension of the measurement protocol to many-body qudit systems and other systems of interest for studying the nature of scrambling would be an interesting task that we leave for future work. We point out that studies of operator size distributions for many-body systems beyond qubits are also scarce, with some exceptions \cite{zhuang2019scrambling,blok2021}. In particular, we note the case where the system of interest interacts with the environment, thus forcing one to probe scrambling and the operator size distribution in the presence of decoherence \cite{schuster2022}. The NOTOC measurement protocol may be extended to open quantum systems using the analysis presented in Sec.~II~B of Ref.~\cite{PhysRevA.106.042429}, from which a detailed analysis of the effect of decoherence may be carried out. Crucially one should revisit the definition of the coarse-grained operator size distribution \eqref{eq:coarsegrainedprobabilitydist} and consider the effect of decoherence on, e.g., the normalization of this distribution. Additionally, the relationship between the fidelity OTOC \eqref{eq:FOTOCidentity} and the NOTOC \eqref{eq:NOTOCdefinition} becomes more complicated for open quantum systems~\cite{PhysRevA.106.042429}.

\section{Acknowledgments}
The authors acknowledge Ivan H. Deutsch, Lorenza Viola, and Andrew Zhao for insightful discussions. This work is supported by a collaboration between the US DOE and other Agencies. This work is based upon work supported by the U.S. Department of Energy, Office of Science, National Quantum Information Science Research Centers, Quantum Systems Accelerator. P.D.B. acknowledges support from the U.S. National Science Foundation through the FRHTP Grant No. PHY-2116246. K.C. acknowledges support from Ministère de l’Économie et de l’Innovation du Québec and the Natural Sciences and Engineering Research Council of Canada.  Work at the University of Strathclyde was supported by AFOSR grant number FA9550-181-1-0064.

\appendix

\section{Finite difference method details for method A}\label{appendix:PDelements_finitedifference}
The $n$th derivative at $x=0$ of the probability-generating function $F(x,t)$ defined in \eqref{eq:PGF} is implemented in the present work as a forward-only finite difference method \cite{NumericalMethodsForOrdinaryDifferentialEquations}, which takes the form \eqref{eq:forwardonlyfinitediff} and is repeated here for convenience:
\begin{align}
    F^{(n)}(x=0,t) =& (\Delta x)^{-n} \sum_{m=0}^{n+a-1} c_m^{(n,a)} F(m\, \Delta x, t) \nonumber\\
    &+ \mathcal{O}(\Delta x^{a}).\label{app:eq:forwardonlyfinitediff}
\end{align}
Here $a$ is the accuracy of the finite difference method, and the step size $\Delta x$ is the distance between the $n+a$ points $\{x = m \Delta x\}_{m=0}^{n+a-1}$ used to approximate the derivative. The forward-only finite-difference coefficients $c_m^{(n,a)}$ are given in table~\ref{table:forwardonlyfinitedifferencecoefficients} for $a=1$ and in table~\ref{table:forwardonlyfinitedifferencecoefficients_a2} for $a=2$, and were obtained using Ref.~\cite{fdcc}. Figure~\ref{fig:derivative_individual} in Sec.~\ref{subsection:PGF_PDelement_extraction} was created using the forward-only finite difference method \eqref{eq:forwardonlyfinitediff} with the coefficients of table~\ref{table:forwardonlyfinitedifferencecoefficients}.

\begin{table}[h]
    \centering
    \begin{tabular}{c||c|c|c|c|c|c|c|c}
        $n$ & $c_0^{(n,1)}$ & $c_1^{(n,1)}$ & $c_2^{(n,1)}$ & $c_3^{(n,1)}$ & $c_4^{(n,1)}$ & $c_5^{(n,1)}$ & $c_6^{(n,1)}$  \\ \hline
        1 & -1 & 1 & -- & -- & -- & -- & -- \\
        2 & 1 & -2 & 1 & -- & -- & -- & -- \\
        3 & -1 & 3 & -3 & 1 & -- & -- & -- \\
        4 & 1 & -4 & 6 & -4 & 1 & -- & -- \\
        5 & -1 & 5 & -10 & 10 & -5 & 1 & -- \\
        6 & 1 & -6 & 15 & -20 & 15 & -6 & 1 \\
    \end{tabular}
    \caption{Forward-only finite difference coefficients $c_m^{(n,a)}$ for derivatives of degree $n\leq 6$, using accuracy $a=1$. The finite size difference coefficients were obtained using Ref.~\cite{fdcc}.}
    \label{table:forwardonlyfinitedifferencecoefficients}
\end{table}

\begin{table}[h]
    \centering
    \begin{tabular}{c||c|c|c|c|c|c|c|c}
        $n$ & $c_0^{(n,2)}$ & $c_1^{(n,2)}$ & $c_2^{(n,2)}$ & $c_3^{(n,2)}$ & $c_4^{(n,2)}$ & $c_5^{(n,2)}$ & $c_6^{(n,2)}$ & $c_7^{(n,2)}$ \\ \hline
        1 & -3/2 & 2 & -1/2 & -- & -- & -- & -- & -- \\
        2 & 2 & -5 & 4 & -1 & -- & -- & -- & -- \\
        3 & -5/2 & 9 & -12 & 7 & -3/2 & -- & -- & -- \\
        4 & 3 & -14 & 26 & -24 & 11 & -2 & -- & -- \\
        5 & -7/2 & 20 & -95/2 & 60 & -85/2 & 16 & -5/2 & -- \\
        6 & 4 & -27 & 78 & -125 & 120 & -69 & 22 & -3 \\
    \end{tabular}
    \caption{Forward-only finite difference coefficients $c_m^{(n,a)}$ for derivatives of degree $n \leq 6$, using accuracy $a=2$. The finite size difference coefficients were obtained using Ref.~\cite{fdcc}.}
    \label{table:forwardonlyfinitedifferencecoefficients_a2}
\end{table}

While in theory one can increase the accuracy $a$ arbitrarily to minimize the error $\mathcal{O}(\Delta x^{a})$ of the approximation for any $\Delta x \ll 1$, in practice the method becomes increasingly sensitive to noise as we increase the accuracy $a$. This increased sensitivity to noise is due to the number of points sampled $n+a$ limiting the values that $\Delta x$ can take, and in the vicinity of vanishing $\Delta x$ the dominant error is not the approximation of the derivative in \eqref{app:eq:forwardonlyfinitediff}, but rather rounding errors due to the presence of the noise \cite{NumericalMethodsForOrdinaryDifferentialEquations}.

\begin{figure}
    \centering
    \includegraphics[width=1\linewidth]{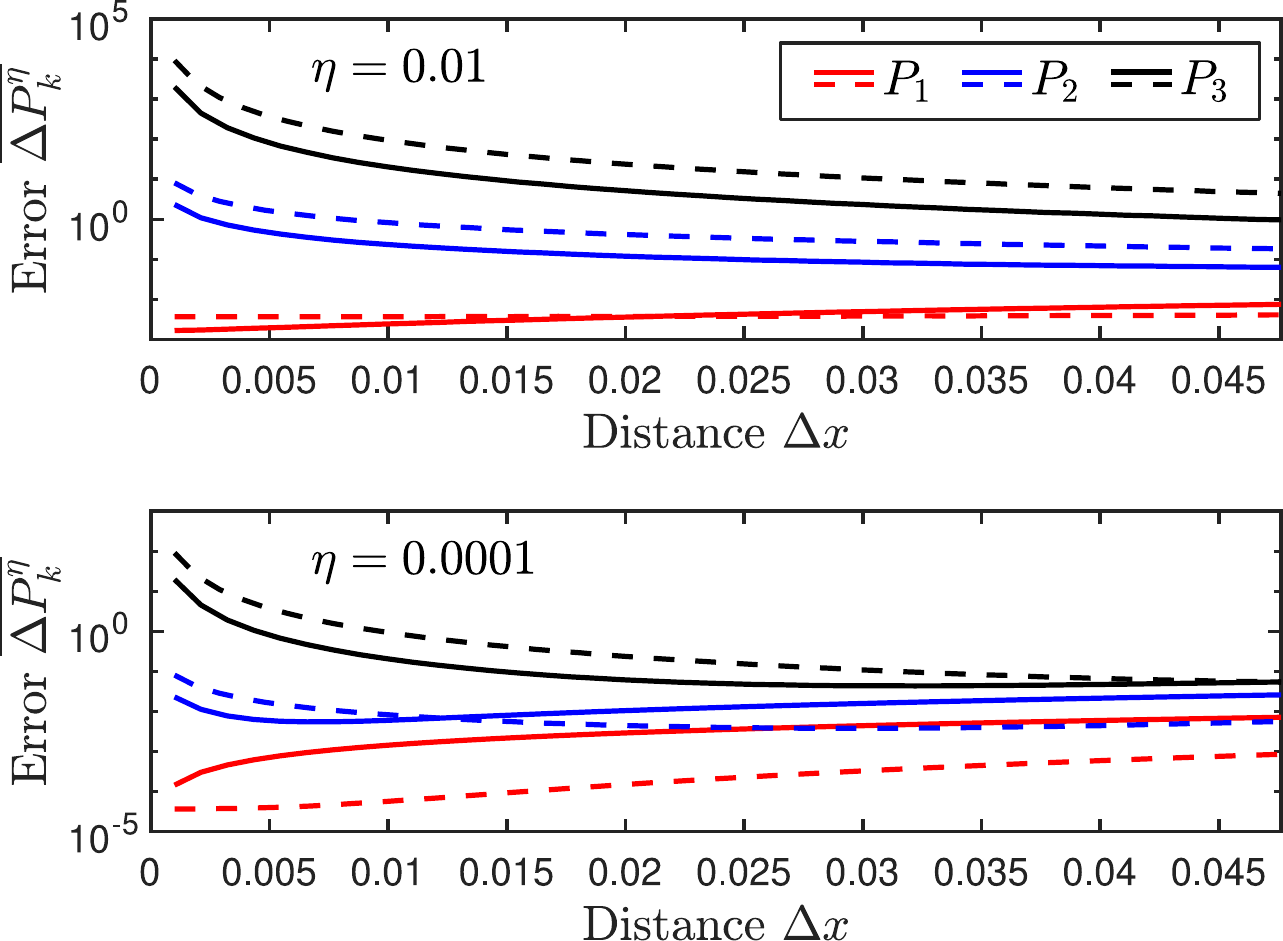}
    \caption{Comparison of time averaged error $\overline{\Delta P_k^\eta}$ [\eqref{eq:TimeAveragedError}] for accuracy $a=1$ (solid curves) and $a=2$ (dashed curves). The used parameters are identical to those used for \fref{fig:derivative_individual}. Shown are the time averaged errors for $P_1$ (red), $P_2$ (blue), and $P_3$ (black curves), for noise amplitude $\eta = 10^{-2}$ (upper panel) and $\eta = 10^{-4}$ (lower panel).}
    \label{app:fig:derivative_comparison}
\end{figure}

Figure~\ref{app:fig:derivative_comparison} displays a comparison between the extraction of the elements of the probability distribution for accuracies $a=1$ (solid curves) and $a=2$ (dashed curves). Plotted is the time averaged error $\overline{\Delta P_k^\eta}$ as defined in \eqref{eq:TimeAveragedError}, averaged over times $Jt \in [0,10]$ for each realization of the noise \eqref{eq:multiplicative_noise} and subsequently averaged over 100 noise realizations. We are considering the same system parameters and operator as in Sec.~\ref{subsection:PGF_PDelement_extraction}. We observe for the higher noise amplitude $\eta = 10^{-2}$ (upper panel) that the lower accuracy method generally outperforms the higher accuracy method for the considered step sizes $\Delta x$, the exception being for $P_1$ when considering $\Delta x \geq 0.02$. The crossover between the $a=1$ and $a=2$ curves at $\Delta x \approx 0.02$ for $P_1$ indicates a transition from being dominated by the sensitivity of the method to the added noise \eqref{eq:multiplicative_noise} for smaller $\Delta x$ values, to being dominated by the error due to the finite difference method's approximations for larger $\Delta x$ values.

For the case of $\eta = 10^{-4}$ (lower panel in \fref{app:fig:derivative_comparison}) we clearly see that the $a=2$ method outperforms the $a=1$ method for $P_1$, and a clear crossover point is also observed for $P_2$. Therefore, when working with a sufficient small noise amplitude such as $\eta = 10^{-4}$, it may be beneficial to use a higher accuracy for the finite difference method.

\section{Reducing state counts in Method B by exploiting symmetries} \label{app:symmetry}
Here we show how to reduce the number of states required to produce in Method B of Sec.~\ref{section:nonepsilonprotocol} by exploiting symmetries. In many cases, we are interested in the operator size distribution of an operator $W$ which is a single site Pauli operator.
For concreteness, let us assume this operator acts on site 1, $W\equiv W_1$. The NOTOC of \eqref{eq:AveragedNOTOC} is constructed by measuring the expectation value of $W_1$ at the end of the protocol. In principle, however, there is no 'cost' associated to measuring expectation values on different sites and using those get more refined information about the operator dynamics. Suppose the system has a translational invariance described by an operator $T_l$
\begin{equation}
    T_l^\dagger \left(\bigotimes\limits_j q_j\right) T_l = \bigotimes\limits_j q_{(j+l \mod{N})},
\end{equation}
such that, for instance, $T_1^\dagger (A\otimes B\otimes C)T_1 = C\otimes A\otimes B$, etc. Consider our system's evolution given by $U(t)$ such that $T_l^\dagger U(t) T_l = U(t)$. Then we have
\begin{align}
    W_1(t) & = \sum\limits_Q f[Q;W_1(t)] Q \\
    T_l^\dagger U(t)^\dagger W_1 U(t) T_l & = \sum\limits_Q f[Q;W_1(t)] T_l^\dagger Q T_l\\
    W_{1+l}(t) & = \sum\limits_Q f\left[T_l Q_l T_l^\dagger,W_1(t)\right] Q_l
\end{align}
when we have defined $Q_l \equiv T_l^\dagger Q T_l$ and used the translation invariance property. By definition, the LHS of the last equation equals
\begin{equation}
    W_{1+l}(t) = \sum\limits_{Q'} f\left[Q',W_{1+l}(t)\right] Q'
\end{equation}
and so we have that
\begin{equation}
    f\left[Q,W_{1+l}(t)\right] = f[T_l Q T_l^\dagger,W_1(t)].
\end{equation}

Suppose we start from a fixed initial state $\rho_1 = \frac{1}{d}\left(\mathbb{I} + \sum_{Q\in C_k}^{'} r_Q Q\right)$, where the sum is over only a subset of operators of weight $k$, starting at position 1 up to $k$. If we measure $\langle W_{1+l}(t)\rangle$ for $l=0,\ldots,N-1$, then
\begin{widetext}
\begin{align}
    \langle W_{1+l}(t)\rangle & = \sum\limits_Q f\left[Q;W_{1+l}\right] \langle Q\rangle_{\rho_1} \\
    & = \sum\limits_Q f\left[T_l Q T_l^\dagger;W_1(t)\right] \mathrm{Tr}\left(\rho_1 Q\right) = \sum\limits_Q f\left[T_l Q T_l^\dagger;W_1(t)\right] \mathrm{Tr}\left(T_l\rho_1 T_l^\dagger T_l Q T_l^\dagger\right) \\
    & = \sum\limits_{\tilde{Q}} f\left[\tilde{Q};W_1(t)\right] \langle \tilde{Q}\rangle_{T_l \rho_1 T_l^\dagger}.
\end{align} 
\end{widetext}

In conclusion, that means that measuring $\langle W_{1+l}(t)\rangle$ yields the same result as having done the protocol starting from $T_{l} \rho_1 T_{l}^\dagger$, thus reducing the state count by a factor of $N$ provided one can measure expectation values in all sites in the case of full translational invariance. The procedure works similarly if one only has a reflection symmetry (i.e. open boundary conditions).

\bibliography{Reference.bib}

\end{document}